\documentclass[12pt]{article}

\usepackage[pdftex]{graphicx}
\usepackage{url}

\usepackage{natbib}
\usepackage{indentfirst}
\usepackage{graphics}
\usepackage{amsmath}
\usepackage{amssymb}
\usepackage{bm}
\usepackage{multirow}
\usepackage{booktabs}
\usepackage{float}
\usepackage{titlesec}
\usepackage{currvita}
\usepackage[applemac]{inputenc}
\usepackage{algorithm}
\usepackage{algpseudocode}
\usepackage{xcolor}
\usepackage{url}
\usepackage{indentfirst}
\usepackage{graphicx}
\usepackage{subcaption}
\usepackage{amsmath}
\usepackage{amssymb}
\usepackage{bm}
\usepackage{multirow}
\usepackage{booktabs}
\usepackage{siunitx}
\usepackage{makecell}
\usepackage{float}
\usepackage{titlesec}
\usepackage{diagbox}
\usepackage{hyperref}
\usepackage[T1]{fontenc}
\usepackage{currvita}
\usepackage[applemac]{inputenc}
\usepackage{tabu}
\usepackage{times}
\usepackage{algorithm, algpseudocode}
\usepackage{amssymb, amsmath, mathrsfs, amsfonts, graphicx, multirow, url}
\usepackage{adjustbox}
\usepackage{array, epsfig, rotating, pdflscape, bm}
\usepackage{algorithm, algpseudocode}
\usepackage{enumerate}
\usepackage{etoolbox}
\usepackage{xcolor}
\usepackage{tikz}
\usepackage{pict2e,picture}
\usetikzlibrary{shapes.symbols, shapes.geometric, arrows, positioning, chains}
\numberwithin{equation}{section}

\def\bSig\mathbf{\Sigma}

\newcommand{\bA} {\mathbf{A}}
\newcommand{\bI} {\mathbf{I}}
\newcommand{\bW} {\mathbf{W}}
\newcommand{\bX} {\mathbf{X}}
\newcommand{\bx} {\mathbf{x}}
\newcommand{\bY} {\mathbf{Y}}
\newcommand{\bZ} {\mathbf{Z}}
\newcommand{\bz} {\mathbf{z}}
\newcommand{\bb} {\mathbf{b}}
\newcommand{\bc} {\mathbf{c}}

\newcommand{\br} {\mathbf{r}}
\newcommand{\bs} {\mathbf{s}}
\newcommand{\bt} {\mathbf{t}}
\newcommand{\bbeta} {\boldsymbol{\beta}}
\newcommand{\bpi} {\boldsymbol{\pi}}
\newcommand{\btheta} {\boldsymbol{\theta}}
\newcommand{\bxi} {\boldsymbol{\xi}}
\newcommand{\bzero}{\mathbf{0}}
\newcommand{\bone} {\mathbf{1}}

\title{Post-selection Inference in Regression Models for Group Testing Data}

\author{
  Qinyan Shen\thanks{\texttt{qshen@email.sc.edu}} \and
  Karl Gregory\thanks{\texttt{gregorkb@stat.sc.edu}} \and
  Xianzheng Huang\thanks{\texttt{huang@stat.sc.edu}} \\
}
\date{} 

\begin{document}

\maketitle

\begin{abstract}
We develop methodology for valid inference after variable selection in logistic regression when the responses are partially observed, that is, when one observes a set of error-prone testing outcomes instead of the true values of the responses. Aiming at selecting important covariates while accounting for missing information in the response data, we apply the expectation-maximization algorithm to compute maximum likelihood estimators subject to LASSO penalization. Subsequent to variable selection, we make inferences on the selected covariate effects by extending post-selection inference methodology based on the polyhedral lemma. Empirical evidence from our extensive simulation study suggests that our post-selection inference results are more reliable than those from naive inference methods that use the same data to perform variable selection and inference without adjusting for variable selection.
\end{abstract}

\section{Introduction}
\label{sec:intro}
When testing a large number of individuals for a disease or infection, it can be advantageous to test individuals in groups rather than one-by-one. This strategy is called group testing.  In group testing, individuals are placed (usually at random) into groups (often small, of 2 to 10 individuals), and the specimens drawn from the individuals in each group are combined into a single pooled specimen. Then the pooled specimen is tested for the presence of the disease or infection, giving a positive or negative result for each group.  If individual diagnoses are desired, individuals in groups which tested positive can be retested one-by-one; depending on the goals of the study, one may or may not retest individuals in groups that tested positive.
\par
One advantage of group testing is that it can greatly reduce the cost of disease surveillance. Even if individuals in positive pools are retested, it is the case that when the disease is rare, the majority of the pools will test negative, resulting in a significant reduction in the total number of tests that need to be performed.  A more hidden advantage to group testing is that, when positive pools are retested, one can, if the disease is rare, obtain more accurate estimators of disease prevalence, etc. than one would obtain if one tested individuals one-by-one. The reason for this is that each test is assumed to be liable to error; that is, it can give a false positive or a false negative result.  The source of the hidden advantage can be put simply: The smaller the number of tests, the fewer the opportunities to make one of these errors.  The sensitivity and specificity of the tests will therefore play an important role in any analysis of group testing data.  The sensitivity of a test is the probability that it is positive when the infection is present, while its specificity is the probability that it is negative when the infection is absent.  Throughout our work we will assume that the sensitivity and specificity of the group or individual tests are known and that they are less than $1$, so that each testing outcome is liable to error.  Given the liability to error of each test, we will throughout this work make a distinction between testing outcomes associated with individuals, which we can observe, and individuals' true disease statuses, which we \textit{cannot} observe.

Group testing has been implemented with success in many settings. \cite{sinnott2020evaluation} demonstrated that the number of individuals it was possible to test was doubled with only a marginal increase in the number of testing kits needed when utilizing group testing for SARS-CoV-2 RNA detection; \cite{verdun2021group} further optimized group testing strategies tailored to the COVID-19 pandemic, leading to a 10-fold increase in testing efficiency. \cite{pilcher2020group} developed a model to assess the efficiency and accuracy of specimen pooling algorithms for SARS-CoV-2, and showed that group testing could enable rapid scale-up of testing and real-time surveillance of incidence. \cite{mcmahan2017bayesian} report that the State Hygienic Laboratory at the University of Iowa, which tests thousands of Iowans annually for chlamydia and gonorrhea, has saved approximately \$3.1 million from 2009 to 2014 by adopting a group testing approach.

Besides disease detection, researchers are interested in performing logistic regression with group testing data in order to identify risk factors or important covariates for the prediction of disease incidence at the individual level. The main challenge to fitting regression models with group testing data is that instead of observing individuals' true disease statuses, we observe testing outcomes which are only guaranteed to be correct with certain probabilities.  As we will see in the development of our methods, the likelihood functions based on the observed testing outcomes are much more complicated than the likelihood functions based on individuals' true disease statuses, and they depend intricately on the sensitivity and specificity of the tests. \cite{xie2001regression},  \cite{zhang2013group}, and \cite{gregory2019adaptive}  addressed these complications by employing the expectation-maximization (EM) algorithm to estimate parameters in  generalized linear regression models for the individual response based on group testing data. \cite{mcmahan2017bayesian} took a Bayesian approach for regression analysis of group testing data.  \cite{joyner2020mixed} formulated a Bayesian generalized linear mixed model that can accommodate any group testing protocol, and used  a spike-and-slab prior to facilitate variable selection. Focusing on multiple-infection group testing data, \citet{lin2019regression} proposed a method exploiting the LASSO penalty for variable selection to identify risk factors for each infection. 

Subsequent to variable selection, one may wish to make inferences on the effects of the selected covariates, such as testing hypotheses regarding covariate effects or constructing confidence intervals for them. A naive approach to inference after variable selection is to fit the regression model using only the selected covariates and then to make inference as though the process of choosing the covariates had not used the same data. There is abundant evidence in the literature that this naive approach can lead to inflated Type I error rates and confidence intervals that are too narrow \citep{berk2013valid, lee2016exact, tibshirani2016exact}. As our own simulation studies demonstrate, naive post-selection inference will have the same undesired consequences in the case of regression with group testing data.  We present a procedure for making valid post-selection inferences in this setting.

The rest of the paper is organized as follows: We provide in Section~\ref{sec:review} a brief review of post-selection inference literature. In Section~\ref{sec:individual}, we present the EM algorithm for LASSO-penalized estimation of regression parameters in a logistic regression model for individual testing data when the individual tests have some sensitivity Se and specificity Sp. We then derive a method for valid post-selection inference. Section~\ref{sec:group} extends the method to group testing data. Section \ref{sec:simu} presents simulation studies of the empirical performance of the proposed methodology and illustrates its implementation on a real data set. Section~\ref{sec:disc} outlines key takeaways and suggestions for future research.

 \section{Overview of post-selection inference}
\label{sec:review}
Throughout this paper we will use the terms \textit{model selection} and \textit{variable selection} interchangeably to mean the process of deciding from among a set of available predictors which ones to keep. Post-selection inference aims to account for the uncertainty introduced by the model selection process, in contrast to classical statistical inference, which assumes the model has been selected or is known before the data are observed. \citet{kuchibhotla2022post} provides a comprehensive review on three strategies for post-selection inference. 

One strategy is sample splitting \citep{wasserman2009high, meinshausen2009p, fithian2014optimal, rinaldo2019bootstrapping}, whereby one splits the observed data into training and testing data sets, using the training set to perform model or variable selection and the testing set to perform inference in the selected model. This strategy avoids ``double dipping'' into the observed data and guarantees inferences unaffected by the selection process.  A downside is the loss in power entailed by using part of the data solely for variable selection.  Inference based on the data splitting approach may also be difficult to replicate, as the model selected as well as the inferences in the selected model may be very sensitive to which observations fall in the training versus the testing set \citep{rasines2022splitting}. 

The work on simultaneous inference by \cite{berk2013valid}, along with subsequent works by \cite{zhang2017simultaneous} and \citet{bachoc2019valid, bachoc2020uniformly}, among many others, contributed to the second strategy of post-selection inference. A common goal of methods under this theme is to control the family-wise error rate for all parameters in the selected model. 
By striving for ``universally valid post-selection inference for all model selection procedures..."  \citep[][Section 4.4]{berk2013valid} and controlling the probability of making an error for {\it any} parameter in a selected model, this strategy can lead to overly conservative confidence intervals.

The third strategy, which we adopt in this paper, attempts to make inferences in the selected model which are conditioned on the event that the selected model was selected.  If one can properly condition on the selection event, one can, for example, construct confidence intervals that maintain nominal coverage rates even when constructed with the same data used in model selection. The key step is to find the conditional sampling distributions of one's estimators given the model selection event.  In linear regression, for example, one may begin by considering the sampling distribution of the least squares estimators (LSEs) in a given, pre-specified model. Then one introduces a selection event, which is the event that the given model is selected by a variable selection procedure.  If the selection event can be expressed as affine constraints on the responses, then analytic derivations of the conditional distributions of the LSEs given the selection event become more feasible.  It has been shown that variable selection in linear regression via forward stepwise selection, least angle regression, and LASSO-penalization results in selection events that place the vector of responses in a polyhedral set \citep{lee2016exact,tibshirani2016exact}.  Due to this property, one finds, in the case of Gaussian noise, that the sampling distributions of the LSEs in the selected model are truncated Gaussians.  These truncated distributions properly account for variable selection and can thus be used to make valid post-selection inferences. \cite{tibshirani2018uniform} showed that this strategy works asymptotically in the case of non-Gaussian noise, and \cite{taylor2018post} outlined a generalization of these results to logistic regression and proportional hazards models. \cite{hyun2018exact} developed inference conditioning on model selection events that are defined by the generalized LASSO regularization path. More recent developments in this direction aim to analytically derive exact or approximately valid post-selection inference under more complicated model settings with different model selection procedures \citep{panigrahi2021integrative, panigrahi2022approximate, rugamer2022post, kuchibhotla2020valid, zhao2022selective, neufeld2022tree}. 


To make valid post-selection inferences in logistic regression with group testing data, we follow the third strategy. We first introduce our covariate effect estimators as well as a selection event. Then we describe how to make post-selection inferences based on the approximate conditional sampling distributions of our estimators given the selection event. We introduce our procedure in the context of logistic regression with individual testing data, in which each individual is tested with a test having some sensitivity Se and some specificity Sp.  Then we extend our methods to the group testing setting in Section \ref{sec:group}.

\section{Post-selection inference based on individual testing data}
\label{sec:individual}
\subsection{The post-selection statistic}
\label{sec:postselest}

Denote by $\bY = (Y_1,\ldots,Y_n)^\top$ a vector of binary responses indicating the unobservable true disease status of $n$  individuals, and by $\bX = (\bx_1,\ldots,\bx_n)^\top$ an $n\times p$ design matrix with values of $p$ covariates for each of the $n$ individuals. We assume a logistic regression model such that $Y_i \mid  \bx_i \sim \mbox{Bernoulli}(\pi_i(\btheta))$ for $i = 1,\dots,n$, with  $\pi_i(\btheta)=1/\{1+\exp(-\alpha - \bx_i^\top\bbeta)\}$, where $\btheta=(\alpha, \bbeta^\top)^\top$. Throughout we will regard $\bx_1,\dots,\bx_n$ as fixed, so we will no longer explicitly condition on the covariate values.

Suppose we observe individual testing outcomes $\bZ=(Z_1, \ldots, Z_n)^\top$ in place of $\bY$, where the tests have sensitivity Se and specificity Sp,  so that $Z_i \mid Y_i \sim \mbox{Bernoulli}(\mbox{Se} \cdot Y_i + (1-\mbox{Sp}) \cdot (1-Y_i))$, for $ i = 1, \ldots, n$. We assume $Z_i|Y_i$ does not depend on $\mathbf{x}_i$. Moreover, we assume Se and Sp to be known. Several works in group testing data make this same assumption, for example \cite{gregory2019adaptive}, \cite{xie2001regression}, and \cite{zhang2013group}, whereas some other works aim to estimate Se and Sp in tandem with the regression coefficients $\btheta$, such as \cite{mcmahan2017bayesian} and \cite{joyner2020mixed}. If treating Se and Sp as unknown, one typically takes a Bayesian approach to estimating them \citep{puggioni2008joint}. In some cases, point estimates of Se and Sp for tests that are in widespread use are available from studies like \cite{haugland2010comparing}, which presented Se and Sp point estimates of $0.890$ and $0.992$, respectively, for a strand displacement assay on cervical swabs for \textit{Chlamydia trachomatis} in women. Given that such information may be available, and so that we may focus on the area of post-selection inference, we will proceed under the assumption that Se and Sp are known.


Viewing $\bY$ as missing data, we obtain the regularized maximum likelihood estimator (MLE) of $\btheta$ using the EM algorithm subject to LASSO penalization, as outlined in Algorithm 1. With complete data $(\bY, \bZ)$, the complete data log-likelihood has the form
\begin{equation}
    \ell_c(\btheta; \bY, \bZ)=\sum_{i=1}^n \left\{ Y_i \log (\pi_i(\btheta)) +(1-Y_i)\log (1-\pi_i(\btheta))\right\}+C(\bY, \bZ),
    \label{eq:completeloglkh}
\end{equation}
where $C(\bY, \bZ)$ is the log-likelihood of $\bZ$  given $\bY$, which is assumed to be free of $\btheta$. In the E-step, we compute the expectation of \eqref{eq:completeloglkh} given $\bZ$ under the current estimate for $\btheta$, say $\btheta^{(k)}$. This gives the objective function (into which we later incorporate the LASSO penalty)
\begin{equation}
     Q(\btheta; \btheta^{(k)}, \bZ)=\sum_{i=1}^n \left\{\mathbb{E}_{\btheta^{(k)}}[ Y_i|Z_i] \log (\pi_i(\btheta)) +(1-\mathbb{E}_{ \btheta^{(k)}}[Y_i|Z_i])\log (1-\pi_i(\btheta))\right\}, 
     \label{eq:Qfun}
\end{equation}
where we have removed the term involving $C(\bY,\bZ)$ since it does not depend on $\btheta$, and where, for $i=1, \ldots, n$, we have
\begin{equation}
\label{eq:E-step}
\mathbb{E}_{\btheta^{(k)}}[Y_i | Z_i ]
=\frac{\text{Se}^{Z_i}(1-\text{Se})^{1-Z_i}\pi_i(\btheta^{(k)})}{\{P_{\btheta^{(k)}}(Z_i=1)\}^{Z_i}\{P_{\btheta^{(k)}}(Z_i=0)\}^{1-Z_i}},
\end{equation}
in which $P_{\btheta^{(k)}}(Z_i=0)=(1-\text{Se})\pi_i(\btheta^{(k)})+\text{Sp}\{1-\pi_i(\btheta^{(k)})\}$, and $P_{\btheta^{(k)}}(Z_i=1)=1-P_{\btheta^{(k)}}(Z_i=0)$. In the M-step, one updates the estimate for $\btheta$ by maximizing  $Q(\btheta; \btheta^{(k)}, \bZ)$ in \eqref{eq:Qfun} penalized by LASSO with a penalty parameter $\lambda \ge 0$. The outputs at convergence of Algorithm 1 are the penalized MLEs for the regression coefficients in $\btheta$. 
\begin{algorithm}[H]
\caption{\label{alg:EM} {\small EM algorithm for LASSO-penalized logistic regression based on imprecise individual testing data}}
  \begin{algorithmic}[1]
        \State Initialize an estimate $\btheta^{(0)}$ for $\btheta$. Set the iteration counter at $k=0$.
        \State Compute $\hat{Y_i}^{(k)}=\mathbb{E}_{\btheta^{(k)}}\left[Y_i| Z_i \right]$, for $i = 1, \ldots, n$, according to \eqref{eq:E-step}.
        \State Obtain 
        \begin{equation}
        \label{eq:update}
        \btheta^{(k+1)} \longleftarrow \underset{\btheta\in \mathbb{R}^{p+1}}{\operatorname{argmax}} \sum_{i=1}^n\left\{\hat Y_i^{(k)} \log \left(\pi_i(\btheta)\right)+(1-\hat Y_i^{(k)}) \log \left(1-\pi_i(\btheta)\right)\right\} - \lambda \|\boldsymbol{\beta}\|_1.
        \end{equation}
        \State Repeat Steps 2 and 3 until $\frac{1}{\sqrt{n}}\|\hat \bY^{(k+1)}-\hat \bY^{(k)}\|_2<10^{-6}$.
  \end{algorithmic}
\end{algorithm}

Incorporating the LASSO penalty into each M-step allows one to perform variable selection. Let $M=\{j\in \{1, \ldots, p\}: \, \hat \beta_j\ne 0\}$ be the set containing the indices of the selected covariates and let $|M|$ be the cardinality of $M$. Subsequent to variable selection, we consider making inferences on $\btheta_M=(\alpha, \bbeta_M^\top)^\top$ based on $\bZ$ and $\bX_M$, where $\bbeta_M$ contains the regression coefficients associated with the selected covariates, and $\bX_M$ is the $n\times |M|$ design matrix with the columns of $\bX$ with indices in $M$. Similarly, let $\hat \btheta_M=(\hat \alpha, \hat \bbeta_M^\top)^\top$. We will base our post-selection inferences on the estimator
\begin{equation}
\bar \btheta_M= (\bar \alpha , \bar \bbeta_M^\top)^\top = \Bigl( [\bone_n, \bX_M]^\top \hat \bW [\bone_n, \bX_M]  \Bigr)^{-1} [\bone_n, \bX_M]^\top \hat \bW \hat{\bz},
\label{eq:WLSE}
\end{equation}
where $\hat \bW=\text{diag}(\hat \pi_1(1-\hat \pi_1), \ldots, \hat \pi_n(1-\hat \pi_n))$, and $\hat \bz = \hat \alpha \bone_n +\bX\hat\bbeta + \hat \bW ^{-1}( \hat{\bY} - \hat\bpi )$, in which $\hat \bY=(\hat Y_1, \ldots, \hat Y_n)^\top$ and $\hat\bpi=(\hat \pi_1, \ldots, \hat \pi_n)^\top$, with $\hat Y_i=\mathbb{E}_{\hat \btheta}[Y_i|Z_i]$ and $\hat \pi_i=\mathbb{E}_{\hat \btheta}[Y_i|\bx_i]$, for $i = 1, \ldots, n$. 
The post-selection estimator in \eqref{eq:WLSE} is motivated by the iteratively reweighted least squares (IRLS) algorithm for fitting the logistic regression model for $\bY$ based on the complete data log-likelihood given in \eqref{eq:completeloglkh} evaluated at $\hat \bY$. In other words, we use the estimated individual responses in $\hat \bY$ at convergence from the EM algorithm as if they were the observed data, and, with $\hat \btheta_M$ as an initial estimate in the IRLS algorithm, we obtain a weighted LSE of $\btheta_M$ that minimizes $(\hat \bz-\alpha-\bX_M \bbeta_M)^\top \hat \bW (\hat \bz-\alpha-\bX_M \bbeta_M)$. Without involving model selection, one may consider making inferences based on the MLE of $\btheta_M$. But this estimator in logistic regression is not available in closed form, which makes it a difficult statistic to use as the first ingredient for developing conditional selective inference. Our choice of the post-selection estimator $\bar \btheta_M$ in \eqref{eq:WLSE} resembles an LSE  in Gaussian linear regression, the setting in which we can establish post-selection inference results.

Next, we characterize the selection event and relate this event to the post-selection estimator to prepare for deriving its conditional distribution given the selection event.

\subsection{The selection event}
\label{sec:selev}
The event on which we will condition in order to make post-selection inference is the event that our LASSO-penalized estimation results in the selection of the covariate set $M$ as well as in the vector of signs given by $\mbox{sign}(\hat \bbeta_M)$ for the estimated coefficients of the selected covariates.  
Following the work in \cite{lee2016exact}, we show next that the selection event can be characterized as a set of affine constraints on the estimator $\bar \bbeta_M$ introduced in \eqref{eq:WLSE}.

The selected model is revealed by $\hat \btheta$ at convergence of the EM algorithm. If one views the estimated individual probabilities of disease $\hat \bY$ at convergence as the actual response data, then the M-step in the last iteration of \eqref{eq:update} is equivalent to the minimization of $-\ell_c(\btheta; \hat \bY, \bZ)$ plus the LASSO penalty. This can be written in weighted least squares form as
 \begin{equation}
 \label{eq:lastMstep}
     \underset{(\alpha, \bbeta)\in \mathbb{R}^{p+1}}{\operatorname{argmin}}(\hat \bz-\alpha-\bX \bbeta)^\top \hat \bW (\hat \bz-\alpha-\bX \bbeta)+\lambda \|\bbeta\|_1. 
 \end{equation} 
According to the Karush-Kuhn-Tucker (KKT) conditions, $\hat \btheta_M=(\hat \alpha, \hat \bbeta_M^\top)^\top$ satisfies
\begin{align}
    \bone_n^\top\hat \bW (\bone_n\hat\alpha + \bX_M\hat\bbeta_M - \hat{\bz}) & = \mathbf{0}, \label{eq:intercept} \\
    \bX_M^\top\hat{\bW}(\bone_n\hat\alpha + \bX_M\hat\bbeta_M - \hat{\bz}) + \lambda \bs_M & = \mathbf{0}, \label{eq:active}\\
    \bX_{-M}^\top\hat{\bW}(\bone_n\hat\alpha + \bX_M\hat\bbeta_M - \hat{\bz}) + \lambda \bs_{-M} & = \mathbf{0},\label{eq:inactive}
\end{align}
where $\bX_{-M}$ is the $n\times (p-|M|)$ matrix containing the columns of $\bX$ with indices not in $M$, $\bs_M = \mbox{sign}(\hat\bbeta_M)$, and $\bs_{-M}$ is a $(p-|M|)\times 1$ vector with each entry in $[-1, 1]$. The equations \eqref{eq:intercept}, \eqref{eq:active}, and \eqref{eq:inactive} are similar to the KKT conditions from \cite{taylor2018post}. 
Setting $\mathcal{J}(\hat\btheta_M)  =  [\bone_n,\bX_M]^\top \hat{\bW} [\bone_n,\bX_M]$, \eqref{eq:intercept} and \eqref{eq:active} allow us to connect the penalized estimator $\hat \bbeta_M$ and the post-selection estimator $\bar \bbeta_M$ with the expression
\begin{equation}
\label{eq:thetabarthetahat}
(\hat \alpha,\hat \bbeta_M^\top)^\top = (\bar \alpha, \bar \bbeta_M^\top)^\top - \mathcal{J}^{-1}(\hat \btheta_M)[0, \lambda \mathbf{s}_M^\top]^\top.
\end{equation}
From here we can translate the selection event to constraints on $\bar \bbeta_M$.


Adapting the formulation of a selection event resulting from least squares estimation subject to the LASSO penalty in \citet[][Theorem 4.3]{lee2016exact} to the penalized weighted least squares problem in \eqref{eq:lastMstep}, the selection event can be expressed as affine constraints on $\bar \bbeta_M$ implied by $\{\text{sign}(\hat \bbeta_M)=\bs_M\}$.  Using \eqref{eq:thetabarthetahat}, one can show that these constraints are equivalent to $\{\mathbf{A}_1\bar\bbeta_M \leq \bb_1\},$
where $\mathbf{A}_1 = -\mbox{diag}(\bs_M)$, and
\begin{equation}
   \mathbf{b}_1 = -\mbox{diag}(\bs_M)
   [\bzero ~ \bI_{|M|}] \mathcal{J}^{-1}(\hat\btheta_M) [0,\lambda \bs_M^\top ]^\top.
   \label{eq:b1}
\end{equation}
We next adjust an approximate pre-selection sampling distribution of $\bar \bbeta_M$ for these constraints, obtaining an approximate post-selection sampling distribution for a linear contrast $\bxi^\top\bar \bbeta_M$ conditional on the selection event.

\subsection{The distribution of the post-selection estimator}
\label{sec:postseldist}
Suppose the truth of $\btheta$ gives $\{j\in \{1, \ldots, p\}: \,  \beta_j\ne 0\} \subseteq M$, that is, the selected model $M$ is correct in the sense that it contains all the truly important covariates. Rearranging \eqref{eq:thetabarthetahat} gives $\bar \btheta_M=\hat \btheta_M +\mathcal{J}^{-1}(\hat \btheta_M)[0, \lambda \bs_M^\top]^\top$, where $\mathcal{J}(\hat \btheta_M)$ can be viewed as the observed Fisher information corresponding to the weighted least squares regression for fitting the selected model, and $[0, \lambda \bs_M^\top]^\top$ as the corresponding score evaluated at $\hat \btheta_M$. Thus, $\bar \btheta_M$ can be viewed as a one-step update from the initial value $\hat \btheta_M$ in the context of an unpenalized weighted least squares regression for a correct model. By Theorem 7.3.3 regarding a one-step estimator in \citet{lehmann1999elements}, we have $\sqrt{n}(\bar \btheta_M-\btheta_M)\stackrel{d}{\to}N(\bzero, \mathcal{I}^{-1}(\btheta_M))$, provided $\hat \btheta_M$ is $\sqrt{n}$-consistent for $\btheta_M$, where $\mathcal{I}(\btheta_M)$ is the Fisher information corresponding to the weighted least squares regression. The $\sqrt{n}$-consistency of $\hat \btheta_M$ as the solution to \eqref{eq:lastMstep} can be achieved by choosing the penalty parameter $\lambda$ such that $\lim_{n\to \infty}\lambda/\sqrt{n}=0$  \citep[Theorem 2, ][]{fu2000asymptotics}.

Hence, the asymptotic sampling distribution of $\bar \btheta_M$ without conditioning on the selection event is $N(\btheta_M, \mathcal{I}^{-1}(\btheta_M)/n)$ if the selected model $M$ contains all truly important covariates and $\lambda/\sqrt{n}\to 0$, while viewing that $\hat \btheta_M$ results from \eqref{eq:lastMstep}, where we ignore the dependence of $\hat \bz$ and $\hat \bW$ on estimates for $\btheta$. Under these assumptions, the asymptotic distribution of $\bar \bbeta_M$ is $N(\bbeta_M, \mathcal{I}^{-1}_M(\btheta_M)/n)$, where $\mathcal{I}^{-1}_M(\btheta_M)$ is the lower $|M|\times |M|$ block of $\mathcal{I}^{-1}(\btheta_M)$. 

To see how the selection event $\{\bA_1 \bar \bbeta_M\le \bb_1\}$ affects the sampling distribution of $\bar \bbeta_M$, we introduce two random quantities that are independent of each other. The first is a linear contrast $\bxi^\top \bar \bbeta_M$ in $\bar \bbeta_M$, where $\bxi\in \mathbb{R}^{|M|}$ is a user-specified vector depending on one's inference target. For example, if one wishes to make inferences on $\beta_j$ for $j \in M$, one sets $\bxi$ as the $j$th elementary basis vector in $\mathbb{R}^{|M|}$. The second quantity is $\br=(\bI_{|M|}-\bc \bxi^\top)\bar \bbeta_M$, where $\bc=\mathcal{I}^{-1}_M(\btheta_M) \bxi\{\bxi^\top \mathcal{I}^{-1}_M(\btheta_M)\bxi\}^{-1}$. 
By Lemma 5.1 in \cite{lee2016exact}, we have $\{\bA_1\bar\bbeta_M \leq \bb_1 \} = \{ v^-(\br) \le \bxi^\top \bar\bbeta_M \le v^+(\br), \,  v^0(\br) \ge 0 \},$
where 
\begin{align*}
    v^-(\br) & = \max_{\{j: \, (\bA_1 \bc)_j < 0\} } \frac{\bb_{1j} - (\bA_1\br)_j}{(\bA_1\bc)_j}, \, \,
    v^+(\br) = \max_{\{j: \, (\bA_1\bc)_j > 0\} } \frac{\bb_{1j} - (\bA_1\br)_j }{(\bA_1\bc)_j}, \text{ and}\\
    v^0(\br) & = \max_{\{j: \, (\bA_1\bc)_j = 0\} } \{\bb_{1j} - (\bA_1\br)_j\}.
 \end{align*}
 Here, for a generic vector $\bt \in \mathbb{R}^{|M|}$,  $\bt_j$ or $(\bt)_j$ denotes its $j$th entry. 
Now, using ``$\stackrel{d}{=}$'' to refer to ``follows the same distribution as,'' we have
\begin{align*}
     \ \bxi^\top \bar \bbeta_M\mid 
     \{\bA_1\bar\bbeta_M \leq \bb_1 \} &\stackrel{d}{=}  \bxi^\top \bar \bbeta_M\mid 
   \{ v^-(\br) \le \bxi^\top \bar\bbeta_M \le v^+(\br), \,  v^0(\br) \ge 0 \} \\
   &\stackrel{d}{=} \bxi^\top \bar \bbeta_M\mid  \{ v^-(\br) \le \bxi^\top \bar\bbeta_M \le v^+(\br)\},
\end{align*}
where the second equality owes to the independence between $\bxi^\top \bar \bbeta_M$ and $\br$. From here we see that the selection event corresponds to a truncation of the sampling distribution of our contrast $\bxi^\top\bar \bbeta_M$. Specifically, the conditional asymptotic distribution of $\bxi^\top \bar \bbeta_M$ given the selection event is the truncated Gaussian distribution $TN(\bxi^\top \bbeta_M, \, \bxi^\top \mathcal{I}^{-1}_M(\btheta_M) \bxi/n, \, v^-(\br), \, v^+(\br))$, where $TN(\mu, \, \sigma^2, \, a, \,b)$ denotes the $N(\mu, \, \sigma^2)$ with support truncated to $[a, b]$.


\subsection{Post-selection inference for linear contrasts of covariate effects}
\label{sec:postselinf}
Denote by $F^{[a,b]}_{\mu, \sigma^2}$ the cumulative distribution function of $TN(\mu, \,\sigma^2, \, a, \,b)$. Then 
$$\left. F_{\bxi^\top \bbeta_M, \, \bxi^\top \mathcal{I}_M^{-1}(\btheta_M) \bxi/n}^{\left[v^-(\br), \, v^+(\br)\right]}\left(\bxi^\top \bar\bbeta_M\right) \right \vert \{\bA_1 \bar\bbeta_M \le \bb_1\} \sim \operatorname{Unif}(0,1)
$$
asymptotically, which reveals an asymptotic pivotal quantity for $\bxi^\top \bbeta_M$. An asymptotic $(1-a)100\%$ confidence interval (CI) for $\bxi^\top \bbeta_M$ is thus given by
\begin{equation}
\left\{ \mu:  \, a/2 \le F_{\mu, \, \bxi^\top \hat{\mathcal{I}}^{-1} \bxi}^{\left[v^-(\hat\br), \, v^+ (\hat \br)\right]}\left(\bxi^\top \bar\bbeta_M\right) \leq 1-a/2\right\}, \label{eq:CI}
\end{equation}
where $\hat \br=(\bI_{|M|}-\hat \bc \bxi^\top)\bar \bbeta_M$, $\hat \bc=\hat{\mathcal{I}}^{-1}\bxi(\bxi^\top \hat{\mathcal{I}}^{-1} \bxi)^{-1}$, and $\hat{\mathcal{I}}$ is an estimate of $n\mathcal{I}_M(\btheta_M)$.


Following Louis' method \citep{louis1982finding}, we construct $\hat{\mathcal{I}}$ based on the observed total information after the last iteration of the EM-algorithm that is given by 
\begin{align}
    \mathcal{I}_{\text{obs}}(\btheta_M; \bZ) & = \mathbb{E}_{\hat \btheta_M}[\mathcal{I}_c(\btheta_M; \bY, \bZ)|\bZ]-\mathbb{E}_{\hat \btheta_M}[\mathcal{S}_c(\btheta_M; \bY, \bZ)\mathcal{S}^\top_c(\btheta_M; \bY, \bZ)|\bZ], \label{eq:obsinfo}
\end{align}
where $\mathcal{S}_c(\btheta_M; \bY, \bZ)$ and $\mathcal{I}_c(\btheta_M; \bY, \bZ)$ are the complete data score and observed information matrix, respectively, evaluated at $\btheta_M$. Recall that $\hat \btheta_M$ is the value of the estimator of $\btheta_M$ at convergence of the EM-algorithm. From the complete data log-likelihood in \eqref{eq:completeloglkh}, we have
\begin{align}
    \mathcal{S}_c(\btheta_M; \bY, \bZ) & =\frac{\partial \ell_c(\btheta_M; \bY, \bZ)}{\partial \btheta^\top_M}=\sum_{i=1}^n \{Y_i-\pi_i(\btheta_M)\}
    \begin{bmatrix}
        1 \\ \bx_{M,i}
    \end{bmatrix}, \label{eq:comdatscore}\\
    \mathcal{I}_c(\btheta_M; \bY, \bZ) &  = - \frac{\partial^2 \ell_c(\btheta_M; \bY, \bZ)}{\partial \btheta_M\partial \btheta^\top_M} = \sum_{i=1}^n \pi_i(\btheta_M)\{1-\pi_i(\btheta_M)\}
    \begin{bmatrix}
        1 & \bx^{\top}_{M,i} \\
        \bx_{M,i} & \bx_{M,i}\bx^{\top}_{M,i}
    \end{bmatrix}, 
    \label{eq:comdatinfo}
\end{align}
where $\bx^\top_{M,i}$ gives the $i$th row of $\bX_M$. Since \eqref{eq:comdatinfo} does not depend on $\bY$, the first expectation in \eqref{eq:obsinfo} is equal to \eqref{eq:comdatinfo}. To derive the second expectation in \eqref{eq:obsinfo}, let $\mathcal{D}(\btheta_M; \bY)=
 \{\bY-\bpi(\btheta_M)\}\{\bY-\bpi(\btheta_M)\}^\top$, where  $\bpi(\btheta_M)=(\pi_1(\btheta_M), \ldots, \pi_n(\btheta_M))^\top$. Then \eqref{eq:comdatscore} implies that 
\begin{equation*}
    \mathcal{S}_c(\btheta_M; \bY, \bZ)\mathcal{S}^\top_c(\btheta_M; \bY, \bZ)= 
    \begin{bmatrix}
        \bone_n^\top\mathcal{D}(\btheta_M; \bY)\bone_n & \bone_n^\top\mathcal{D}(\btheta_M; \bY)\bX_M \\
        \bX_M^\top\mathcal{D}(\btheta_M; \bY)\bone_n & 
        \bX_M^\top\mathcal{D}(\btheta_M; \bY)\bX_M
    \end{bmatrix}.
\end{equation*}
 Hence, the second term in \eqref{eq:obsinfo} depends on $\bZ$ only via $\mathbb{E}_{\hat \btheta_M}[\mathcal{D}(\btheta_M; \bY)|\bZ]$, of which the $[i, k]$ entry is, for $i\ne 
 k\in \{1, \ldots, n\}$, 
\begin{equation}  
\mathbb{E}_{\hat \btheta_M}[\{Y_i-\pi_i(\btheta_M)\}\{Y_k-\pi_k(\btheta_M)\}|\bZ]  
= \{\hat Y_i-\pi_i(\btheta_M)\}\{\hat Y_k-\pi_k(\btheta_M)\},
 \label{eq:Eik} 
 \end{equation}
 and, for $i=k\in \{1, \ldots, n\}$, 
 \begin{equation}  
\mathbb{E}_{\hat \btheta_M}[\{Y_i-\pi_i(\btheta_M)\}^2|\bZ]  
= \{1-2\pi_i(\btheta_M)\}\hat Y_i+\pi^2_i(\btheta_M), 
 \label{eq:Eikeq} 
 \end{equation}
 where $\hat Y_i=\mathbb{E}_{\hat \btheta_M}[Y_i|Z_i]$ is given by \eqref{eq:E-step} (with $\btheta^{(k)}$ there set at $\hat \btheta_M$), for $i=1, \ldots, n$. We then use the lower $|M|\times |M|$ block of $\mathcal{I}_{\text{obs}}(\hat \btheta_M; \bZ)$ as an estimate of $n\mathcal{I}_M(\btheta_M)$ in \eqref{eq:CI}, that is, 
 \begin{equation}
     \hat{\mathcal{I}}=\sum_{i=1}^n \pi_i(\hat \btheta_M)\{1-\pi_i(\hat \btheta_M)\} \bx_{M,i}\bx^\top_{M,i}-\bX_M^\top \mathbb{E}_{\hat \btheta_M}[\mathcal{D}(\hat\btheta_M; \bY)|\bZ] \bX_M. \label{eq:IMhat}
 \end{equation}

\section{Adaptation for group testing data}
\label{sec:group}
We now adapt the strategy of Section~\ref{sec:individual} to group testing data. Suppose $n$ individuals are randomly partitioned into $J$ pools, and let $\mathcal{P}_1,\dots,\mathcal{P}_J$ be the partition of $\{1,\dots,n\}$ such that $\mathcal{P}_j$ is the set of indices of the individuals in pool $j$. Set $Y_j^* = \max _{i \in \mathcal{P}_j} Y_i$, which is the true disease status for pool $j$ (equal to 1 if at least one individual is positive), and let $Z_j$ be the observed testing result for the pool, where the test has sensitivity Se and specificity Sp.


In this setting, the complete data log-likelihood will still be of the form in \eqref{eq:completeloglkh}, so Algorithm 1 can still be used to compute the LASSO-penalized MLE: The only change is in how we compute the conditional expectations of the unobserved true disease statuses $Y_1,\dots,Y_n$ given the observed  testing outcomes, which are now the pooled testing outcomes $Z_1,\dots,Z_J$.  As long as we can compute these conditional expectations, we can compute $\bar \btheta_M$ and follow the arguments in Section $\ref{sec:individual}$ for making valid post-selection inferences on contrasts in $\bar \bbeta_M$.


Under group testing, for $i \in  \mathcal{P}_j$, we have
\begin{equation}
\label{eq:Estepgp}
\mathbb{E}_{\btheta}[Y_i | Z_j ]
=\frac{\text{Se}^{Z_j}(1-\text{Se})^{1-Z_j}\pi_i(\btheta)}{\{P_{\btheta}(Z_j=1)\}^{Z_j}\{P_{\btheta}(Z_j=0)\}^{1-Z_j}},
\end{equation}
where $P_{\btheta}(Z_j = 0)=(1-\mbox{Se})[1-\prod_{i \in \mathcal{P}_j}  \{1-\pi_i (\btheta)\}] + \mbox{Sp}\prod_{i \in \mathcal{P}_j}  \{1-\pi_i(\btheta)\}$, and $
P_{\btheta}(Z_j = 1) =1-P_{\btheta}(Z_j = 0)$. Now, the LASSO-penalized maximum likelihood estimator is computed with Algorithm 1 as in the individual testing case, but with $\hat Y_i^{(k)}$ on line 2 computed according to the conditional expectation in \eqref{eq:Estepgp}. The output of the EM algorithm gives the penalized MLE $\hat \btheta$ of $\btheta$, which reveals a selected model $M$, under which we obtain the estimated responses in $\hat \bY$, $\hat Y_i=\mathbb{E}_{\hat \btheta}[Y_i|Z_j]$,  for $i\in \mathcal{P}_j$ and $j=1, \ldots, J$, according to \eqref{eq:Estepgp}. Using $\hat \bY$ and $\hat \btheta$, we define the post-selection estimator $\bar \btheta_M$ as in \eqref{eq:WLSE}. 


Having obtained $\bar \btheta_M$, we follow the steps in Section \ref{sec:individual} until we come to the estimation of the observed total information $n \mathcal{I}_M(\btheta_M)$. The identity in \eqref{eq:obsinfo} on which Louis' method is based still holds, only now $\bZ$ represents the group testing outcomes. As a result, the estimated total information associated with $\bar \bbeta_M$ is still given by \eqref{eq:IMhat}, but with the entries in $\mathbb{E}_{\hat \btheta_M}[\mathcal{D}(\btheta_M; \bY)|\bZ]$ derived differently, as elaborated in Web Appendix A. The $[i,k]$ entry of $\mathbb{E}_{\hat \btheta_M}[\mathcal{D}(\btheta_M; \bY)|\bZ]$ is now found as follows: For $i,k \in \mathcal{P}_j$ with $i \neq k$, we have
\begin{align*}
   &\ \mathbb{E}_{\hat \btheta_M}[\{Y_i-\pi_i(\btheta_M)\}\{Y_k-\pi_k(\btheta_M)\}|\bZ] \\
  =&\ \frac{\text{Se}^{Z_j}(1-\text{Se})^{1-Z_j}\pi_i(\hat\btheta_M)\pi_k(\hat\btheta_M)}{\{P_{\hat\btheta_M}(Z_j=1)\}^{Z_j}\{P_{\hat\btheta_M}(Z_j=0)\}^{1-Z_j}}- \pi_k(\btheta_M)\hat Y_i-\pi_i(\btheta_M)\hat Y_k +\pi_i(\btheta_M)\pi_k(\btheta_M),
\end{align*}
where $\hat Y_i=\mathbb{E}_{\hat \btheta_M}[Y_i|Z_j]$ and $\hat Y_k=\mathbb{E}_{\hat \btheta_M}[Y_k|Z_j]$ are given by \eqref{eq:Estepgp} with $\hat \btheta_M$ replacing $\btheta$; for $i\in \mathcal{P}_j$ and $k\in \mathcal{P}_m$, $j\ne m$, the entry is still given by \eqref{eq:Eik}, but with $\hat Y_i=\mathbb{E}_{\hat \btheta_M}[Y_i|Z_j]$ and $\hat Y_k=\mathbb{E}_{\hat \btheta_M}[Y_k|Z_m]$; if $i = k \in \mathcal{P}_j$, the entry is also given by \eqref{eq:Eikeq}, but with the estimated response $\hat Y_i$ obtained based on the group response $Z_j$ according to \eqref{eq:Estepgp}.

If $M$ is an incorrect model, i.e., $M$ does not contain all truly active covariates, then we view $\btheta_M$ as the ``truth'' defined by an equation similar to Equation (1.2) in \cite{lee2016exact}. Then all discussions in Sections 3 and 4 still go through, but, in the presence of model misspecification, one would use the sandwich-form of variance-covariance matrix estimation in \cite{elashoff2004algorithm} to estimate the variance-covariance instead of Louis' method. This estimator, which is derived in Web Appendix B, gives some protection against model misspecification in variance estimation. 

Our method can be easily extended to accommodate more complicated group testing schemes, for example, when individuals in positive pools are retested. While more complicated testing schemes induce more complicated likelihood functions, one can still implement the EM algorithm described in Algorithm 1 and use Louis' method for evaluating the observed information matrix: One needs only to be able to compute conditional expectations of the true, unobserved disease statuses $Y_1,\dots,Y_n$ given all the observed testing outcomes.  Suppose $\mathcal{A}_i$ represents all the testing outcomes which play a role in the conditional probability that $Y_i = 1$. For example, if groups of individuals are tested in pools and then individuals in positive pools are re-tested, $\mathcal{A}_i$ will consist of a single negative test if the pool in which individual $i$ was placed tests negative; on the other hand, if the pool in which individual $i$ was placed tests positive, $\mathcal{A}_i$ will consist of the positive pool test as well as the re-testing outcomes for all individuals in the pool.  To implement our methods in this setting, we can replace the condition expectations given in equations \eqref{eq:E-step} or \eqref{eq:Estepgp} by $\mathbb{E}_{\boldsymbol{\theta}}[Y_i|\mathcal{A}_i]$. See, for example, \cite{gregory2019adaptive} for details on how to compute the conditional expectations $\mathbb{E}_{\boldsymbol{\theta}}[Y_i|\mathcal{A}_i]$ for some more complicated testing schemes.

\section{Empirical evidence}
\label{sec:simu}
We shall acknowledge that several approximations are made in developing the proposed post-selection inference procedure in Sections \ref{sec:individual} and \ref{sec:group}. First, in \eqref{eq:lastMstep}, we view $\hat \bz$ as response data and $\hat \bW$ as the weight matrix even though they depend on an estimate of $\btheta$. The uncertainty in this estimate when formulating the selection event is ignored, for instance, by viewing $\bb_1$ in  \eqref{eq:b1} as fixed given a selected model even though it depends on $\hat \btheta_M$. Second, we approximate the pre-selection sampling distribution of $\bar \btheta_M$ by a Gaussian distribution. Effects of these approximations permeate through the post-selection distribution of $\bxi^\top \bar \bbeta_M$ and the CI in \eqref{eq:CI}. Despite being only approximately valid post-selection inference procedures, they are expected to improve over their naive counterparts which assume the selection event to be known and fixed when making inferences on $\bbeta_M$. This claim is supportd by empirical evidence from our extensive simulation study. 

\subsection{Simulation design}
\label{sec:simdes}
In our simulation study, we generate true individual responses $\{Y_i\}_{i=1}^n$ from a logistic regression model with $p=10$ covariates and regression coefficients $\btheta=(-5, 2, 1, 1, \bzero^\top)^\top$, that is, $\alpha=-5$ and only the first three covariates are truly important. Realizations of each of the ten covariates are generated independently from the $N(0, 1)$ distribution. We then randomly form $J$ pools such that $|\mathcal{P}_j| = m$ for all $j$ for some pool size $m$, and compute true pool responses $\{Y^*_j=\max_{i\in \mathcal{P}_j} Y_i\}_{j=1}^J$. Lastly, we simulate the imprecise group testing response $Z_j$ from $\text{Bernoulli}(\text{Se} \cdot Y_j^*+(1-\text{Sp})(1-Y_j^*))$, for $j=1, \ldots, J$,  with $\text{Se}=0.95$ and $\text{Sp}=0.97$. We consider the sample sizes $n\in \{1000, \, 2000\}$ and the pool sizes $m \in \{1, \, 2, \, 4\}$, where $m=1$ leads to $\{Z_j\}_{j=1}^J$ as imprecise individual testing data. A total of $5000$ Monte Carlo replicates are generated at each combination of $n$ and $m$. 

We set the penalty parameter $\lambda$ at a sequence of values ranging from 1 to 7 (when $n=1000$) or 0.5 to 10 (when $n=2000$) that are equally spaced on the logarithmic scale. Using each Monte Carlo replicate dataset, we fit the logistic regression model following Algorithm 1 at each pre-specified $\lambda$; then we carry out the proposed post-selection inference to construct 95\% CI's for each entry of $\bbeta_M$, where $M$ is the selected model. We also obtain 95\% CI's naively in the classical way assuming the selected model to be known a priori. 

For each simulated dataset, we compute a realization of the Type I error rate in the following way: After model selection, we build a confidence interval for each entry of $\bbeta_M$, where $M$ is the selected model. If the selected model includes covariates for which the regression coefficient is equal to zero, then for each of these, we check whether the confidence interval for the regression coefficient contains zero. The realized Type I error rate is the proportion of these confidence intervals which did not contain zero. If the selected model $M$ does not include any ``extra'' covariates (covariates with regression coefficient equal to zero), then the realized Type I error rate is recorded as zero for that dataset. Specifically, we compute on each simulated dataset the quantity
$$
\text{Type I error rate} = \frac{|\{j\in M:  \, 0 \notin \text{CI}(\hat\beta_j) \text{ and } \beta_j=0 \}|}{|\{j\in M: \, \beta_j=0\}|},
$$
where $\text{CI}(\hat \beta_j)$ is a $95\%$ confidence interval for $\beta_j$ according to the considered inference procedure, and we record zero if $|\{j\in M: \, \beta_j=0\}|=0$. In addition, we record for each data set, at each preset value of $\lambda$, the Akaike information criterion (AIC) and Bayesian information criterion (BIC) associated with the selected model, that is based on the observed data log-likelihood evaluated at $\hat \btheta_M$.

\subsection{Simulation results}
\begin{figure}[H]
\begin{tabular}{cc}
  \includegraphics[width=80mm]{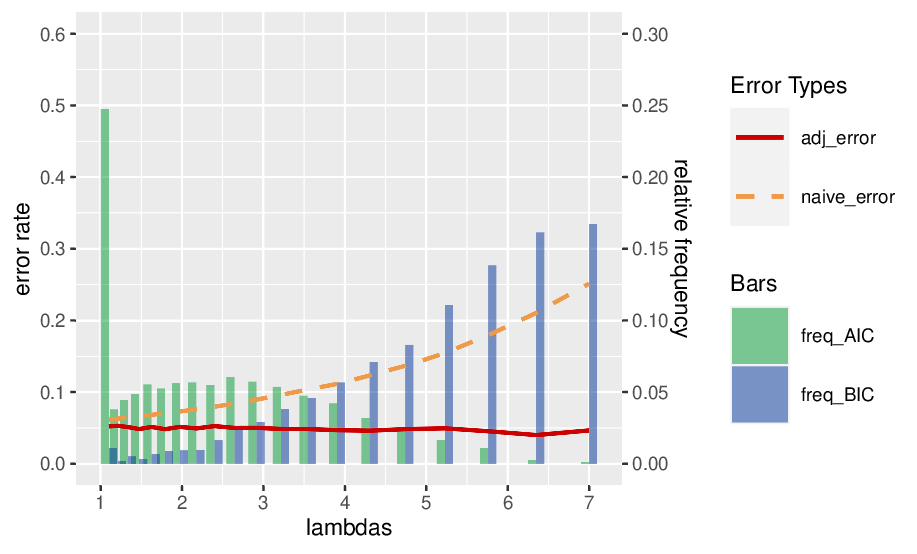} &   \includegraphics[width=80mm]{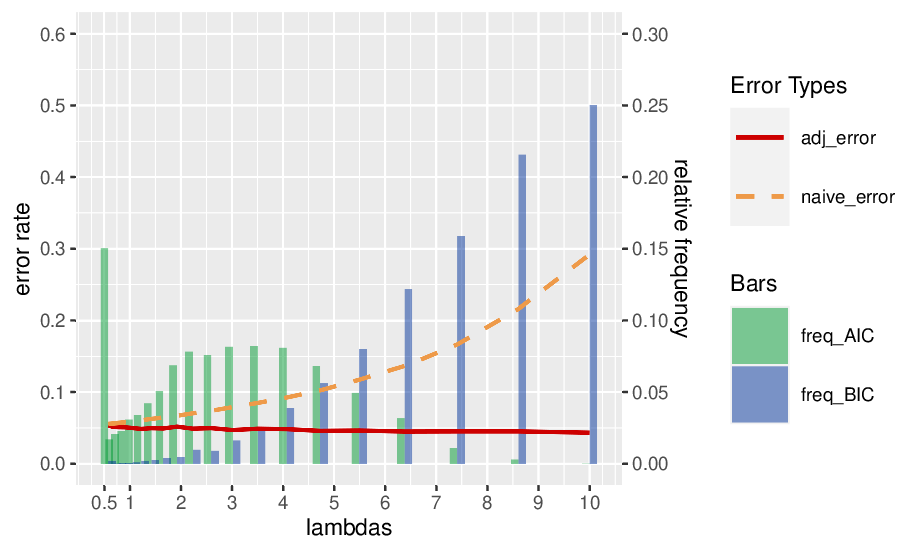} \\
(a) $m = 1, n = 1000$ & (b) $m = 1, n = 2000$ \\[6pt]
 \includegraphics[width=80mm]{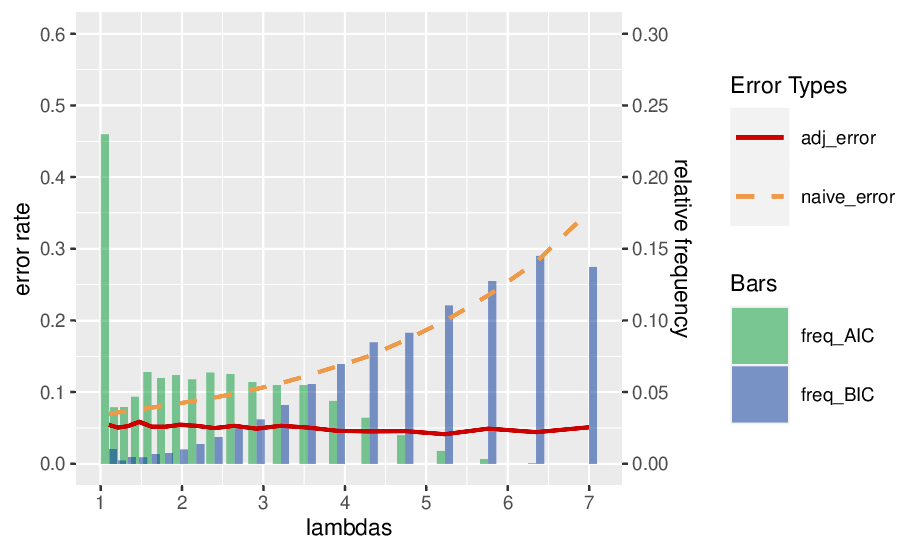} &   \includegraphics[width=80mm]{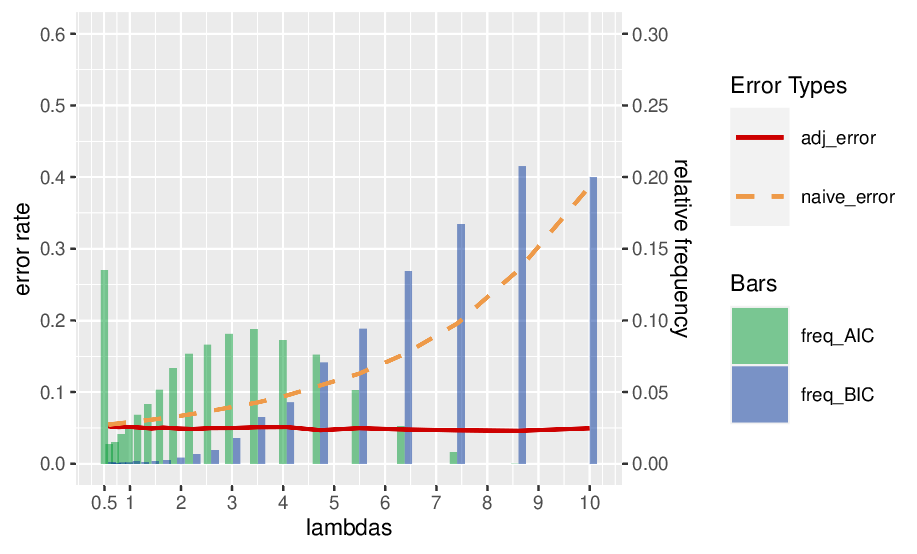} \\
(c) $m = 2, n = 1000$ & (d) $m = 2, n = 2000$ \\[6pt]
 \includegraphics[width=80mm]{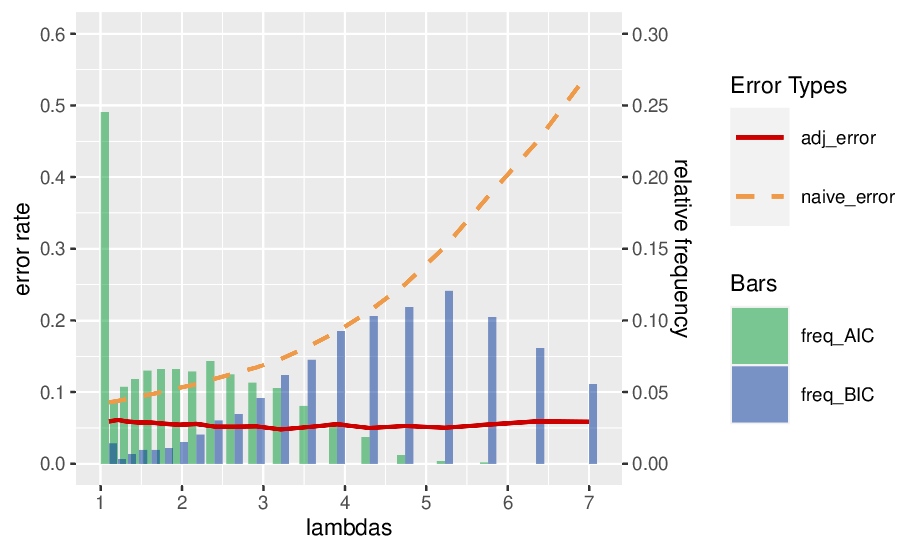} &   \includegraphics[width=80mm]{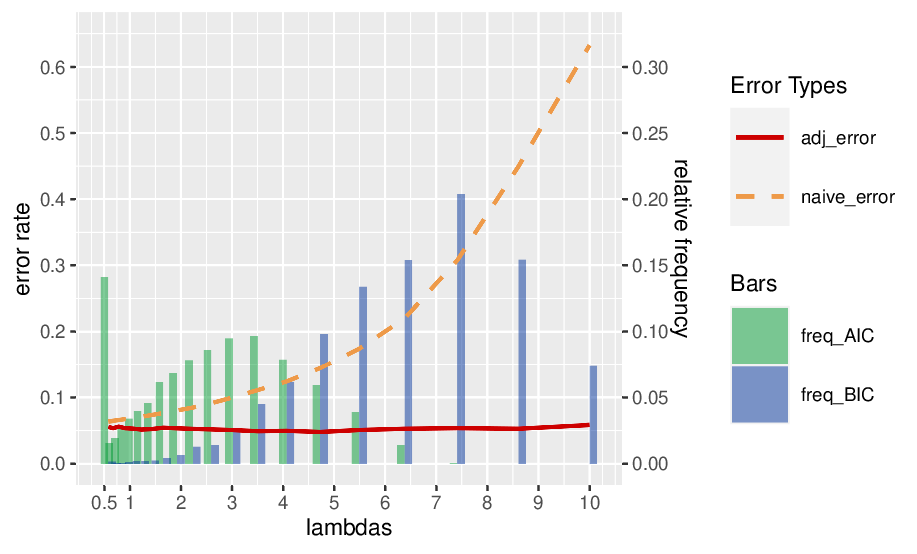} \\
(e) $m = 4, n = 1000$ & (f) $m = 4, n = 2000$ \\\\[6pt]
\end{tabular}
\caption{\label{fig:type1}Averages of Type I error rates across 5000 Monte Carlo replicates versus $\lambda$ when the proposed post-selection inference procedure is used to construct $95\%$ CI's for selected covariate effects (red solid lines), and when the naive inference procedure is used (orange dashed lines) as the sample size $n$ and pool size $m$ vary. Overlaying the plot of Type I error rate versus $\lambda$ are the histogram of chosen $\lambda$ according to AIC (green bars), and the counterpart histogram when BIC is used to choose $\lambda$ (blue bars).}
\end{figure}

Figure~\ref{fig:type1} depicts Monte Carlo averages of the realized Type I error rate associated with the naive and selection-adjusted inference procedures as $\lambda$ increases. In all six simulation settings, our proposed method produces Type I error rates closely matching the nominal level 0.05 across all pre-set values of $\lambda$, whereas the naive method yields an increasingly inflated Type I error as $\lambda$ increases. A larger penalty $\lambda$ makes the regularization more aggressive and typically results in a smaller model size $|M|$. Ignoring the uncertainty in variable selection when inferring selected covariate effects does more harm when one more proactively penalizes large models, and having group testing data with a larger group size only exacerbates the problem. In contrast, our method adequately accounts for variable selection and provides more reliable inference on selected covariate effects based on either individual testing data or group testing data with arbitrary group sizes. 

If one first uses AIC or BIC to choose a value of $\lambda$ and then selects a model using the chosen $\lambda$, the plots in Figure~\ref{fig:type1} suggest that the naive inference will be liable to above-nominal Type I error rates; this problem could be worse under BIC selection than under AIC selection, as the BIC tends to select smaller models (larger values of $\lambda$).  In contrast, the proposed post-selection inference procedure exhibits good control of the Type I error rate over the entire range of $\lambda$ values typically chosen by the AIC or BIC.

Table~\ref{table: results} summarizes the performance of $95\%$ CI's for $\beta_2$, $\beta_4$, and $\beta_6$ resulting from the proposed method and from the naive method along with those for the corresponding odds ratios $\exp(\beta_j)$, $j=2,4,6$, where the odds ratio CIs are constructed by exponentiating the endpoints of the coefficient CIs. Note that $\beta_2$ is nonzero, whereas $\beta_4$ and $\beta_6$ are both equal to zero in the true model. We report for each coefficient the average of the upper as well as the lower CI bound, the average widths of the CI, and the proportions of the CIs that contain the true value of the coefficient---when the corresponding covariate is selected---from 5000 simulated datasets. Results are reported at a single, fixed value of $\lambda$ chosen as follows: For each of the 5000 data sets, we recorded the value of $\lambda$ which minimized the AIC criterion across a grid of $\lambda$ choices (where the same grid of choices was considered for each dataset). Then we set our fixed $\lambda$ value equal to the median of these values. This was done at each combination of $n$ and $m$. Table~\ref{table: results} shows that the naive CI's are on average much narrower than our selection-adjusted CI's constructed according to \eqref{eq:CI}. This explains the inflated Type I error of the naive method exhibited in Figure~\ref{fig:type1}. For $\beta_4$ and $\beta_6$, which are equal to zero, the coverage probabilities of the naive CI's are considerably lower than the nominal level, while those of the adjusted CI's are close to the nominal level. For $\beta_2$, which is nonzero, we see that the coverage probability of the proposed method falls a little below the nominal level; this owes to the shrinkage towards zero of the one-step estimators of the covariate effects, which is not unexpected.  For coefficients equal to zero, the shrinkage towards zero of the one-step estimator does not adversely affect the coverage of the selection-adjusted CI's, since the shrinkage makes the one-step estimator more accurate for zero-valued coefficients.

\begin{table}[H]
\centering
\small
\caption{\label{table: results}Monte Carlo averages of empirical coverage probabilities (C.P.) of $95\%$ CI's  for $\beta_2$, $\beta_4$ and $\beta_6$, along with the corresponding average confidence interval limits, average confidence interval for odds ratio, average widths (A.W.), and average parameter estimates (P.E.). The results are based on a single value of $\lambda$ (the median AIC choice of $\lambda$) across 5000 Monte Carlo simulations at each combination of $n$ and $m$. Se=0.95, Sp=0.97. }
\resizebox{\columnwidth}{!}{
\begin{tabular}{cllccccccccccc}
\hline
\multirow{2}{*}{$n$} & \multicolumn{1}{c}{\multirow{2}{*}{param}} & \multicolumn{1}{c}{\multirow{2}{*}{$m$}} & \multicolumn{5}{c}{Proposed}                            &  & \multicolumn{5}{c}{Naive}                               \\ \cline{4-14} 
                     & \multicolumn{1}{c}{}                       & \multicolumn{1}{c}{}                     & CI                       & CI odds           & A.W. & C.P. & P.E.   &  & CI                       & CI odds           & A.W. & C.P. & P.E.  \\ \hline
                     &                                            & 1                                        & {[}0.28,1.82{]}          & {[}1.32,6.17{]}   & 1.5  & 93.1 & 0.93  &  & {[}0.53,1.68{]}          & {[}1.70,5.37{]}   & 1.2  & 95.6 & 0.83 \\
                     & $\beta_2$                                  & 2                                        & {[}0.19,1.93{]}          & {[}1.21,6.89{]}   & 1.7  & 92.4 & 0.93  &  & {[}0.51,1.70{]}          & {[}1.67,5.47{]}   & 1.2  & 94.7 & 0.83 \\
                     &                                            & 4                                        & {[}0.02,2.16{]}          & {[}1.02,8.67{]}   & 2.1  & 93.9 & 0.93  &  & {[}0.47,1.81{]}          & {[}1.60,6.11{]}   & 1.3  & 95.3 & 0.84 \\ \cline{2-14} 
                     &                                            & 1                                        & {[}-0.89,1.57{]}         & {[}0.41,4.81{]}   & 2.5  & 94.9 & 0.00     &  & {[}-0.46,0.45{]}         & {[}0.63,1.57{]}   & 0.9  & 92.8 & 0.00    \\
1000                 & $\beta_4$                                  & 2                                        & {[}-1.02,1.81{]}         & {[}0.36,6.11{]}   & 2.8  & 94.1 & 0.00     &  & {[}-0.50,0.49{]}         & {[}0.61,1.63{]}   & 1.0  & 91.2 & 0.00    \\
                     &                                            & 4                                        & {[}-1.19,2.33{]}         & {[}0.30,10.29{]}  & 3.5  & 94.2 & 0.00     &  & {[}-0.56,0.58{]}         & {[}0.57,1.79{]}   & 1.1  & 90.1 & 0.00    \\ \cline{2-14} 
\multicolumn{1}{l}{} &                                            & 1                                        & {[}-0.88,1.56{]}         & {[}0.41,4.75{]}   & 2.4  & 95.0 & -0.01 &  & {[}-0.47,0.45{]}         & {[}0.62,1.57{]}   & 0.9  & 92.5 & 0.00    \\
\multicolumn{1}{l}{} & $\beta_6$                                  & 2                                        & {[}-0.97,1.91{]}         & {[}0.38,6.75{]}   & 2.9  & 94.2 & -0.01 &  & {[}-0.50,0.48{]}         & {[}0.61,1.62{]}   & 1.0  & 91.6 & 0.00    \\
\multicolumn{1}{l}{} &                                            & 4                                        & {[}-1.19,2.26{]}         & {[}0.30,9.59{]}   & 3.5  & 94.0 & 0.00     &  & {[}-0.57,0.57{]}         & {[}0.57,1.77{]}   & 1.1  & 89.8 & 0.00    \\ \hline
\multicolumn{1}{l}{} &                                            & 1                                        & {[}0.53,1.53{]}          & {[}1.70,4.62{]}   & 1.0  & 92.9 & 0.96  &  & {[}0.67,1.41{]}          & {[}1.96,4.10{]}   & 0.7  & 94.9 & 0.90 \\
                     & $\beta_2$                                  & 2                                        & {[}0.48,1.57{]}          & {[}1.62,4.81{]}   & 1.1  & 93.3 & 0.95  &  & {[}0.66,1.43{]}          & {[}1.93,4.18{]}   & 0.8  & 94.9 & 0.88 \\
                     &                                            & 4                                        & {[}0.37,1.71{]}          & {[}1.45,5.53{]}   & 1.3  & 93.9 & 0.95  &  & {[}0.63,1.49{]}          & {[}1.88,4.44{]}   & 0.9  & 94.6 & 0.89 \\ \cline{2-14} 
                     &                                            & 1                                        & {[}-0.60,1.06{]}         & {[}0.55,2.89{]}   & 1.7  & 94.8 & 0.00     &  & {[}-0.30,0.31{]}         & {[}0.74,1.36{]}   & 0.6  & 92.8 & 0.00    \\
2000                 & $\beta_4$                                  & 2                                        & {[}-0.67,1.21{]}         & {[}0.51,3.36{]}   & 1.9  & 95.5 & 0.00     &  & {[}-0.32,0.33{]}         & {[}0.72,1.39{]}   & 0.7  & 92.4 & 0.00    \\
                     &                                            & 4                                        & {[}-0.79,1.48{]}         & {[}0.45,4.39{]}   & 2.3  & 94.3 & 0.00     &  & {[}-0.37,0.38{]}         & {[}0.69,1.46{]}   & 0.8  & 91.3 & 0.00    \\ \cline{2-14} 
\multicolumn{1}{l}{} &                                            & 1                                        & {[}-0.60,1.08{]}         & {[}0.55,2.94{]}   & 1.7 & 94.7 & 0.00     &  & {[}-0.31,0.30{]}         & {[}0.73,1.35{]}   & 0.6 & 92.6 & 0.00    \\
\multicolumn{1}{l}{} & $\beta_6$                                  & 2                                        & {[}-0.66,1.22{]}         & {[}0.52,3.39{]}   & 1.9 & 95.2 & 0.00     &  & {[}-0.33,0.32{]}         & {[}0.72,1.38{]}   & 0.7 & 92.2 & 0.00    \\
\multicolumn{1}{l}{} &                                            & 4                                        & {[}-0.80,1.42{]}         & {[}0.45,4.13{]}   & 2.2 & 94.6 & 0.00     &  & {[}-0.37,0.38{]}         & {[}0.69,1.46{]}   & 0.8 & 91.1 & 0.00    \\ \hline
\end{tabular}}
\end{table}

\begin{figure}[H]
\begin{tabular}{cc}
  \includegraphics[width=80mm]{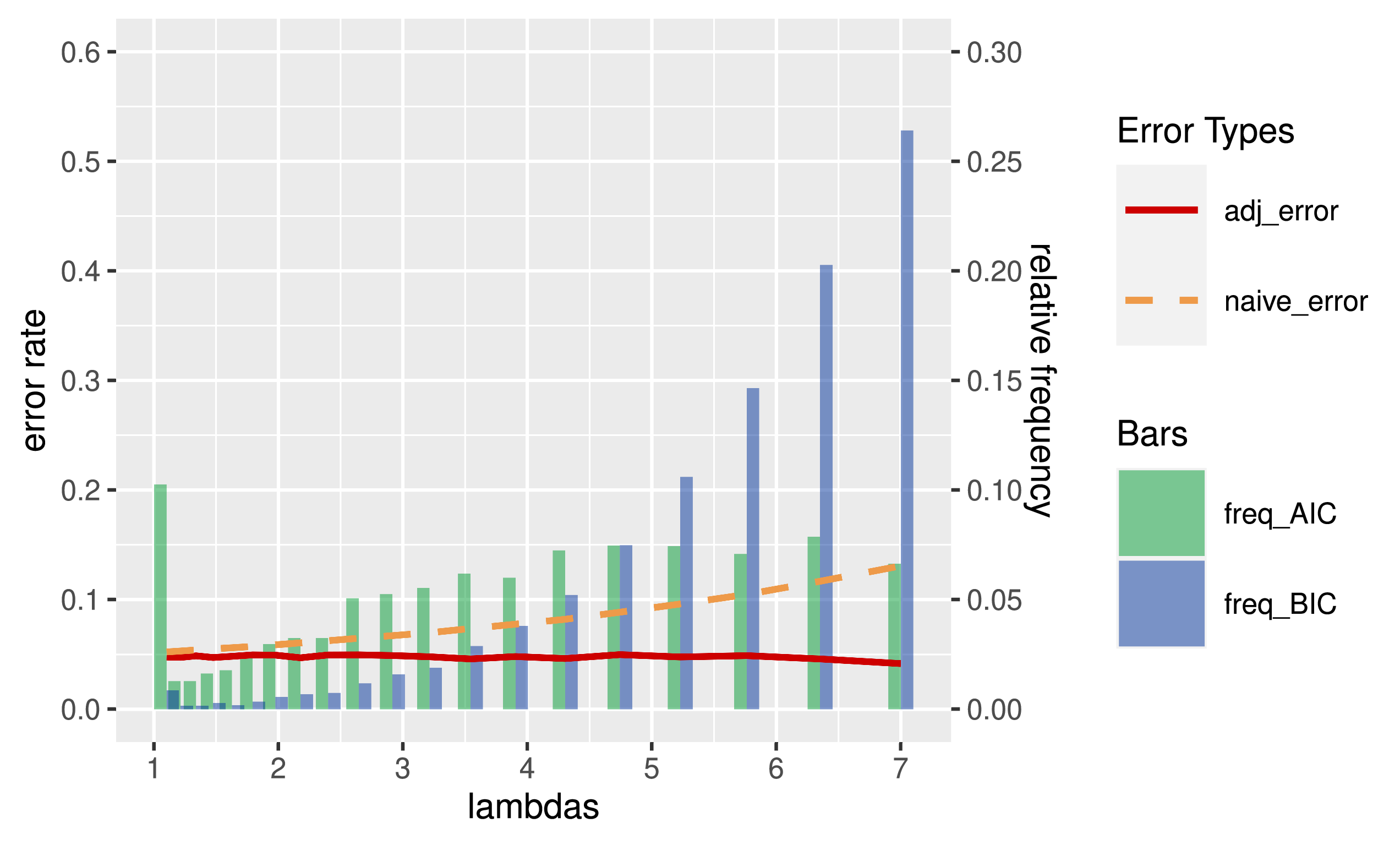} &   \includegraphics[width=80mm]{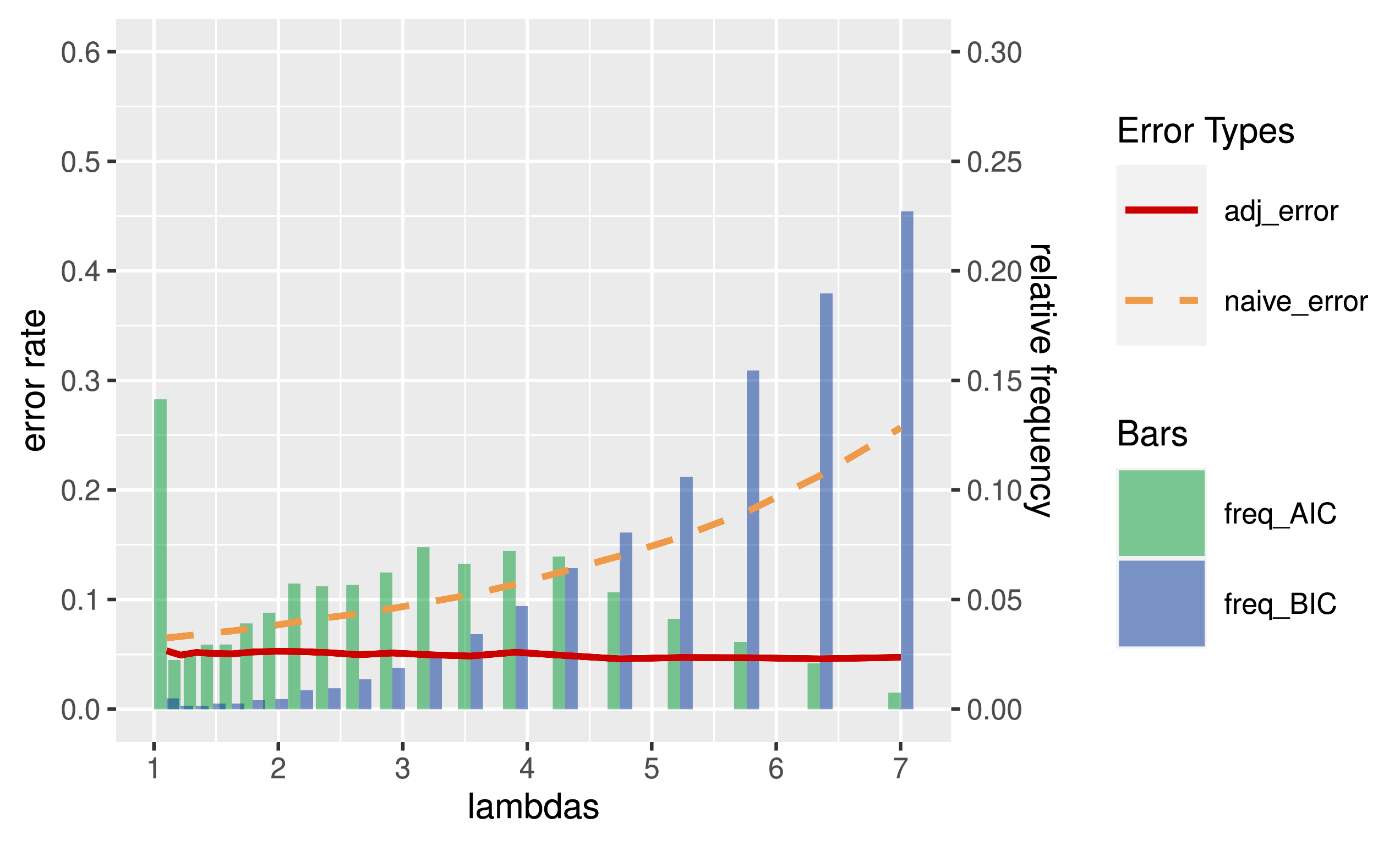} \\
(a) $m = 1, n = 1000$ & (b) $m = 2, n = 1000$ \\\\[6pt]
\end{tabular}
\caption{\label{fig:type1 misspecification}Averages of Type I error rates across 5000 Monte Carlo replicates versus $\lambda$ when the proposed post-selection inference procedure is used to construct $95\%$ CI's for selected covariate effects (red solid lines), and when the naive inference procedure is used (orange dashed lines) as the sample size $n$ and pool size $m$ vary, with the sensitivity and specificity of tests misspecified. Overlaying the plot of Type I error rate versus $\lambda$ are the histogram of chosen $\lambda$ according to AIC (green bars), and the counterpart histogram when BIC is used to choose $\lambda$ (blue bars).}
\end{figure}

\begin{table}[H]
\centering
\small
\caption{\label{table: misspecification results}Monte Carlo averages of empirical coverage probabilities (C.P.) of $95\%$ CI's for $\beta_2$, $\beta_4$ and $\beta_6$, along with the corresponding average confidence interval limits, average confidence interval for odds ratio, average widths (A.W.), and average parameter estimates (P.E.). The results are based on a single value of $\lambda$ (the median AIC choice of $\lambda$) across 5000 Monte Carlo simulations at each combination of $n$ and $m$. Misspecified Se=0.90, Sp=0.92 when generating data. }
\resizebox{\columnwidth}{!}{
\begin{tabular}{cllccccccccccc}
\hline
\multirow{2}{*}{$n$} & \multicolumn{1}{c}{\multirow{2}{*}{param}} & \multicolumn{1}{c}{\multirow{2}{*}{$m$}} & \multicolumn{5}{c}{Proposed}                            &  & \multicolumn{5}{c}{Naive}                               \\ \cline{4-14} 
                     & \multicolumn{1}{c}{}                       & \multicolumn{1}{c}{}                     & CI                       & CI odds           & A.W. & C.P. & P.E.   &  & CI                       & CI odds           & A.W. & C.P. & P.E.  \\ \hline
                     &                        & 1                     & {[}-0.01,0.86{]} & {[}0.99,2.36{]} & 0.9 & 13.8 & 0.40  &  & {[}0.15,0.75{]} & {[}1.16,2.12{]} & 0.6 & 10.5 & 0.34  \\
                     & $\beta_2$              & 2                     & {[}-0.06,1.36{]} & {[}0.94,3.90{]} & 1.4 & 53.4 & 0.57  &  & {[}0.25,1.09{]} & {[}1.28,2.97{]} & 0.8 & 60.7 & 0.51  \\
                     &                        & 4                     & {[}-0.22,1.90{]} & {[}0.80,6.69{]} & 2.1 & 84.0 & 0.72  &  & {[}0.31,1.44{]} & {[}1.36,4.22{]} & 1.1 & 90.0 & 0.65  \\ \cline{2-14} 
                     &                        & 1                     & {[}-0.57,1.01{]} & {[}0.57,2.75{]} & 1.6 & 95.3 & 0.00     &  & {[}-0.28,0.28{]} & {[}0.75,1.32{]} & 0.6 & 92.8 & 0.00     \\
1000                 & $\beta_4$              & 2                     & {[}-0.82,1.52{]} & {[}0.44,4.58{]} & 2.3 & 94.9 & 0.00     &  & {[}-0.39,0.38{]} & {[}0.68,1.46{]} & 0.8 & 90.8 & 0.00     \\
                     &                        & 4                     & {[}-1.15,2.05{]} & {[}0.32,7.77{]} & 3.2 & 93.9 & 0.00     &  & {[}-0.51,0.52{]} & {[}0.60,1.68{]} & 1.0 & 89.3 & 0.00     \\ \cline{2-14} 
\multicolumn{1}{l}{} &                        & 1                     & {[}-0.57,1.00{]} & {[}0.57,2.72{]} & 1.6 & 95.7 & 0.00     &  & {[}-0.28,0.27{]} & {[}0.75,1.31{]} & 0.5 & 92.3 & 0.00     \\
\multicolumn{1}{l}{} & $\beta_6$              & 2                     & {[}-0.83,1.50{]} & {[}0.44,4.48{]} & 2.3 & 94.6 & 0.00     &  & {[}-0.38,0.39{]} & {[}0.68,1.48{]} & 0.8 & 90.6 & 0.00     \\
\multicolumn{1}{l}{} &                        & 4                     & {[}-1.12,2.17{]} & {[}0.33,8.76{]} & 3.3 & 93.6 & -0.01 &  & {[}-0.53,0.50{]} & {[}0.59,1.65{]} & 1.0 & 89.0 & -0.01 \\ \hline
\end{tabular}}
\end{table}

A practical consideration is how our method for post-selected adjustment will perform under misspecification of the sensitivity and specificity of the tests. Table~\ref{table: misspecification results} and Figure~\ref{fig:type1 misspecification} summarize the performance of naive and selection-adjusted inference when the true sensitivity and specificity of the tests are Se = 0.90 and Sp = 0.92, but when the higher values Se = 0.95 and Sp = 0.97 are assumed by the estimation procedure.  This corresponds to the situation in which we assume the tests are more accurate than they really are. According to Figure~\ref{fig:type1 misspecification}, which focuses on Type I error rates, this kind of misspecification does not appear to degrade the performance of our selection-adjusted inference, as the patterns we see are similar to those in Figure~\ref{fig:type1}. From Table~\ref{table: misspecification results}, however, we see that for $\beta_2$, which is nonzero, the coverage probabilities of both the naive and the selection-adjusted CI's are nowhere close to the nominal level.  This shows that misspecifying the sensitivity and specificity of the tests can severely distort both naive and selection-adjusted inferences for nonzero coefficients; this is not very surprising, as this kind of misspecification bases estimation on an incorrect likelihood function.  Somewhat surprising, however, is that Type I error rates (inference on coefficients that are truly equal to zero) appear to be largely unaffected by the misspecification.

\subsection{Application to chlamydia data}

We now analyze data collected through the National Health and Nutrition Examination Survey from year 2015 to 2016 on $n=1371$ participants aged 14 to 39 years. Besides individual testing results regarding chlamydia infection based on urine specimens, participants' demographic information were also collected, such as ethnicity, income level, and age. For illustration purposes, we consider a logistic regression model for the indicator of chlamydia infection, with covariates Age, non-Hispanic white, non-Hispanic black, Mexican American and Hispanic, family monthly poverty level index, new partner in the past 12 months, an indicator of poverty index $>$ 5, gender, education level and marriage status (married or never married). Variable selection was performed in order to identify the most significant risk factors, thereby improving the model performance and interpretability in assessing the likelihood of chlamydia infection. A sensitivity of 0.97 and a specificity of 0.99 are assumed for the test for the diagnosis of chlamydia, which we take to be reasonable values according to \cite{whellams2021screening}.  

\begin{figure}[H]
\begin{center}
\includegraphics[width=150mm]{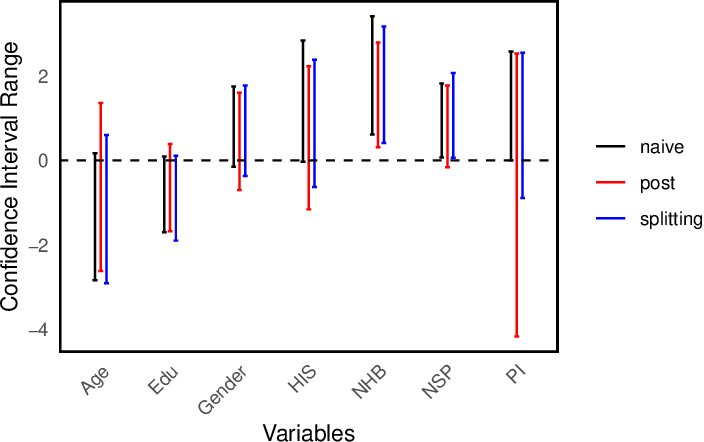} 
\end{center}
\caption{\label{fig: CI_ind}Naive, selection-adjusted, and data-splitting confidence intervals for the regression coefficients of selected covariates under group size $m = 1$ based on the chlamydia data.}
\end{figure}

We first use the individual testing data to fit the logistic regression model subject to LASSO penalty, with $\lambda$ chosen by AIC, followed by constructing $95\%$ CI's for covariate effects in the selected model. As in the simulation study, we compare CI's from the naive method to those from our proposed method. To introduce another competitor, we present results from the strategy of data splitting, using half the data for variable selection and half for interval estimation. Web Appendix C contains results from this strategy implemented in the simulation setting described in Section~\ref{sec:simdes}. Table~\ref{table:realdata_cipe} (the upper half) presents these interval estimates for a selected model resulting from these three approaches. Figure~\ref{fig: CI_ind} provides a pictorial comparison of these confidence intervals. 

\begin{figure}[H]
\begin{center}
\includegraphics[width=150mm]{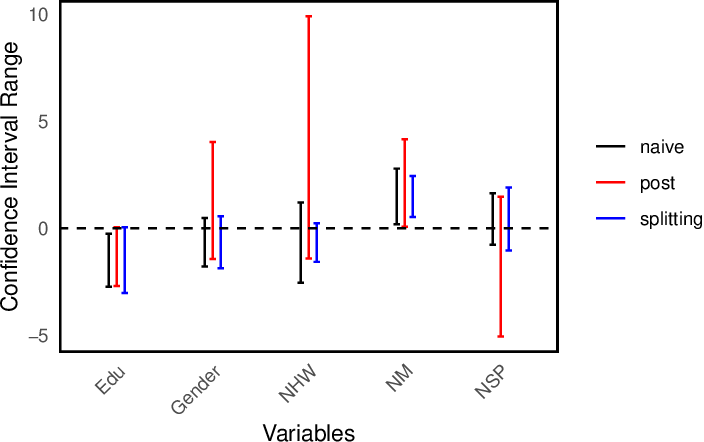} 
\end{center}
\caption{\label{fig: CI_group}Naive, selection-adjusted, and data-splitting confidence intervals for the regression coefficients of selected covariates under group size $m = 3$ based on  the chlamydia data.}
\end{figure}

Finally, we randomly partition the $1371$ individuals into $J=457$ groups of size $m=3$ and artificially generate group testing responses. We repeat the exercise of model selection followed by interval estimation using the aforementioned three strategies. The resulting 95\% CI's are provided in Table~\ref{table:realdata_cipe} (the lower half), with a visual contrast given in Figure~\ref{fig: CI_group}.

\begin{table}[H]
\centering
\small
\setlength\tabcolsep{4.5pt}
\caption{Confidence intervals and point estimates for selected covariate effects from three considered methods based on the chlamydia data}
\label{table:realdata_cipe}
\begin{tabular}{llcccccc}
\hline
\multirow{2}{*}{\textbf{$m$}} & \multirow{2}{*}{Covariate} & \multicolumn{2}{c}{Proposed}           & \multicolumn{2}{c}{Naive}                & \multicolumn{2}{c}{Data Splitting}     \\ \cline{3-8} 
                              &                            & CI                             & PE    & CI                               & PE    & CI                             & PE    \\ \hline
1                             & Age                        & {[}-2.62, 1.36{]}              & -1.22 & {[}-2.84, 0.17{]}                & -1.34 & {[}-2.91, 0.60{]}              & -1.16 \\
                              & Non-Hispanic Black         & {[}0.31, 2.79{]}$\ast$         & 1.72  & {[}0.61, 3.41{]}$\ast$           & 2.01  & {[}0.41, 3.17{]}$\ast$         & 1.79  \\
                              & New Sex Partner            & {[}-0.16, 1.77{]}              & 0.92  & {[}0.07, 1.82{]}$\ast$           & 0.95  & {[}0.06, 2.07{]}$\ast$         & 1.06  \\
                              & Hispanic                   & {[}-1.16, 2.23{]}              & 1.07  & {[}-0.03, 2.84{]}                & 1.41  & {[}-0.63, 2.38{]}              & 0.87  \\
                              & Poverty Index $> 5$        & {[}-4.17, 2.53{]}              & 1.06  & {[}0, 2.58{]}                    & 1.29  & {[}-0.89, 2.55{]}              & 0.83  \\
                              & Gender                     & {[}-0.70, 1.60{]}              & 0.72  & {[}-0.15, 1.75{]}                & 0.80  & {[}-0.37, 1.77{]}              & 0.70  \\
                              & Education Level            & {[}-1.68, 0.39{]}              & -0.82 & {[}-1.70, 0.09{]}                & -0.80 & {[}-1.90, 0.11{]}              & -0.90 \\ \hline
3                             & Non-Hispanic White         & {[}-1.41, 9.91{]}              & -0.31 & {[}-2.55, 1.21{]}                & -0.67 & {[}-1.57, 0.24{]}              & 1.05  \\
                              & New Sex Partner            & {[}-5.06, 1.48{]}              & 0.36  & {[}-0.77, 1.64{]}                & 0.43  & {[}-1.04, 1.91{]}              & 1.01  \\
                              & Gender                     & {[}-1.44, 4.03{]}              & -0.42 & {[}-1.79, 0.48{]}                & -0.65 & {[}-1.87, 0.56{]}              & -2.04 \\
                              & Education Level            & {[}-2.70, 0.05{]}              & -1.37 & {[}-2.73, -0.25{]}$\ast$         & -1.49 & {[}-3.03, 0.05{]}              & -0.37 \\
                              & Never Married              & {[}0.07, 4.16{]}$\ast$         & 1.50  & {[}0.19, 2.79{]}$\ast$           & 1.49  & {[}0.53, 2.45{]}$\ast$         & 0.83  \\ \hline
\end{tabular}
\end{table}

\begin{table}[H]
\centering
\small
\setlength\tabcolsep{4.5pt}
\caption{Odds ratio confidence intervals and point estimates for selected covariate effects from three considered methods based on the chlamydia data}
\label{table:realdata_cipe_odds}
\begin{tabular}{llcccccc}
\hline
\multirow{2}{*}{\textbf{$m$}} & \multirow{2}{*}{Covariate} & \multicolumn{2}{c}{Proposed}           & \multicolumn{2}{c}{Naive}                & \multicolumn{2}{c}{Data Splitting}     \\ \cline{3-8} 
                              &                            & CI                             & PE    & CI                               & PE    & CI                             & PE    \\ \hline
1                             & Age                        & {[}0.07, 3.90{]}             & 0.30 & {[}0.06, 1.18{]}               & 0.26 & {[}0.06, 1.82{]}             & 0.31 \\
                              & Non-Hispanic Black         & {[}1.36, 16.33{]}$\ast$      & 5.59 & {[}1.84, 30.31{]}$\ast$        & 7.46 & {[}1.51, 23.82{]}$\ast$      & 5.99 \\
                              & New Sex Partner            & {[}0.85, 5.87{]}             & 2.51 & {[}1.07, 6.17{]}$\ast$         & 2.59 & {[}1.06, 7.93{]}$\ast$       & 2.89 \\
                              & Hispanic                   & {[}0.31, 9.26{]}             & 2.92 & {[}0.97, 17.13{]}              & 4.10 & {[}0.54, 10.81{]}            & 2.39 \\
                              & Poverty Index $> 5$        & {[}0.02, 12.56{]}            & 2.89 & {[}1.00, 13.22{]}              & 3.63 & {[}0.41, 12.83{]}            & 2.30 \\
                              & Gender                     & {[}0.50, 4.95{]}             & 2.05 & {[}0.86, 5.75{]}               & 2.23 & {[}0.69, 5.87{]}             & 2.01 \\
                              & Education Level            & {[}0.19, 1.48{]}             & 0.44 & {[}0.18, 1.09{]}               & 0.45 & {[}0.15, 1.12{]}             & 0.41 \\ \hline
3                             & Non-Hispanic White         & {[}0.24, 20135.96{]}         & 0.73 & {[}0.08, 3.36{]}               & 0.51 & {[}0.21, 1.27{]}             & 2.86 \\
                              & New Sex Partner            & {[}0.01, 4.39{]}             & 1.43 & {[}0.46, 5.16{]}               & 1.54 & {[}0.35, 6.77{]}             & 2.75 \\
                              & Gender                     & {[}0.24, 56.36{]}            & 0.66 & {[}0.17, 1.62{]}               & 0.52 & {[}0.15, 1.75{]}             & 0.13 \\
                              & Education Level            & {[}0.07, 1.05{]}             & 0.25 & {[}0.07, 0.78{]}$\ast$         & 0.23 & {[}0.05, 1.05{]}             & 0.69 \\
                              & Never Married              & {[}1.07, 64.01{]}$\ast$      & 4.48 & {[}1.21, 16.34{]}$\ast$        & 4.43 & {[}1.70, 11.60{]}$\ast$      & 2.30 \\ \hline
\end{tabular}
\end{table}


From Table~\ref{table:realdata_cipe},  we see that the selection-adjusted method tends to be most conservative when making claims regarding statistical significance of a covariate effect, which is reflected in the count of CI's that exclude zero, as well as in the width of a CI clearly depicted in Figures~\ref{fig: CI_ind} and \ref{fig: CI_group}. More specifically, when $m = 1$, the variable Non-Hispanic Black is deemed significant by all three methods. The variable New Sex Partner is deemed significant by the naive method and data splitting method, but not according to the selected-adjusted method.  Under $m = 3$, data splitting and the selection-adjusted method agree with the naive method on the significance of the variable Never Married, with the widest CI resulting from the selection-adjusted method.  The naive method finds Education Level to be significant, but neither the proposed selection-adjusted method nor data splitting finds it significant.  While data-splitting is a sound method for obtaining confidence intervals and $p$-values which will maintain nominal coverage probabilities and Type I error rates, it is important to note that in the case of highly unbalanced binary response data (when the responses are mostly zeros with a small number of ones or vice versa), the set of selected covariates can vary wildly with the subset of the data chosen to perform variable selection. That is, different random splits of the data can result in selected models which are very different.  In fact, in order to make the comparison between our method and the naive method with the data splitting method, we needed to try many different splits of the data until we found a split under which the selected model was the same as the selected model from the full data set. These considerations pose considerable drawbacks to the data splitting method, in particular when one wants reproducible results. Lastly, Table 4 presents the odds ratio confidence intervals and point estimates for the selected variables. This table is provided for readers who are interested in these additional details.

\section{Discussion}
\label{sec:disc}
We develop in this study valid post-selection inference procedures based on binary response data that are partially observed due to testing error and/or pooling. The strategy of relating the penalized estimation and post-selection estimation of regression coefficients to a weighted least squares problem with Gaussian-like response data (i.e., $\hat \bz$) was also used in 
\citet{taylor2018post} to develop post-selection inference for LASSO-penalized likelihood models. This strategy allows us to draw parallels between the current setting with a partially observed non-Gaussian response and the setting with a linear regression model for a Gaussian response, so that we can borrow established results in the latter setting \citep{lee2016exact, tibshirani2016exact} to derive valid post-selection confidence intervals for target parameters. Although we focus on logistic regression for a binary response here, we believe this strategy can be generalized to other non-Gaussian responses modeled by other generalized linear regression models. The fact that individuals' true response values are unobservable, either due to testing error or pooling, may need to be addressed case-by-case. We use the EM algorithm to address this problem here; other imputation-based methods can be viable alternatives. 

Despite the approximations (in formulating response data and sampling distributions) we invoke in the development of our post-selection inference procedure, their impact on the resulting inference is not noticeable in our simulation study, especially when the penalty parameter $\lambda$ is pre-set at a level such that the LASSO regularization is not too aggressive. In the event of more aggressive penalization, say, by using a penalty chosen by BIC, the proposed method can have the tendency of over-adjusting for model selection, leading to overly conservative confidence intervals that are too wide and produce Type I error much lower than the nominal level. An interesting follow-up research problem is the development of valid post(-double)-selection inference procedures that adequately account for both the selection of $\lambda$ and model selection given the selected $\lambda$.

\vspace*{-8pt}

\bibliographystyle{apalike}

\vspace*{0.5in}

\section*{Supplementary Material}
Web Appendices referenced in Sections~\ref{sec:group} and \ref{sec:simu} available with this paper at the Biometrics website on Oxford Academic. An R package for implementing the proposed methods is available at  \url{https://github.com/kateshen28/Post-LASSO-Inference-Group-Testing/}.

\appendix

\label{lastpage}

\end{document}


\date{}

\pagerange{\pageref{firstpage}--\pageref{lastpage}} 
\volume{}
\pubyear{}
\artmonth{}

\doi{}

\label{firstpage}

\begin{abstract}
The supplementary materials include detailed derivations of quantities referenced in Section 4 and additional simulation results referenced in Section 5 of the main article. 
\end{abstract}

\maketitle
\setcounter{equation}{0} 
\def\theequation{A.\arabic{equation}}
\setcounter{figure}{0} 
\def\thefigure{A.\arabic{figure}}
\section*{Web Appendix A: Derivations of $\mathbb{E}_{\hat \btheta_M}[\mathcal{D}(\btheta_M; \bY)|\bZ]$}
\label{sec:A}
As defined in Section 3 in the main article, $\mathcal{D}(\btheta_M; \bY)=\{\bY-\bpi(\btheta_M)\}\{\bY-\bpi(\btheta_M)\}^\top$. For the $[i,i]$ entry of $\mathcal{D}(\btheta_M; \bY)$, its expectation given $\bZ$ is, assuming $i\in \mathcal{P}_j$, 
\begin{align*}
   \mathbb{E}_{\hat \btheta_M}[\{Y_i-\pi_i(\btheta_M)\}^2|\bZ] & = \mathbb{E}_{\hat \btheta_M}[Y_i^2-2\pi_i(\btheta_M)Y_i+\pi^2_i(\btheta_M)|Z_j]\\ 
   & = \{1-2\pi_i(\btheta_M)\} \mathbb{E}_{\hat \btheta_M}[Y_i|Z_j]+\pi^2_i(\btheta_M), \text{ since $Y_i^2=Y_i$ for a binary $Y_i$,}\\
    & = \{1-2\pi_i(\btheta_M)\} \hat Y_i +\pi^2_i(\btheta_M), 
\end{align*}
where we define $\hat Y_i=\mathbb{E}_{\hat \btheta_M}[Y_i|Z_j]$. 
For the $[i, k]$ entry of $\mathcal{D}(\btheta_M; \bY)$, its expectation given $\bZ$ is, assuming $i\in \mathcal{P}_j$, $k\in \mathcal{P}_m$, and $j\ne m \in \{1, \ldots, J\}$, 
\begin{align*}
    &\ \mathbb{E}_{\hat \btheta_M}[\{Y_i-\pi_i(\btheta_M)\}\{Y_k-\pi_k(\btheta_M)\}|\bZ] \\
     = &\  \mathbb{E}_{\hat \btheta_M}[Y_i-\pi_i(\btheta_M)|Z_j]\mathbb{E}[Y_k-\pi_k(\btheta_M)|Z_m], \text{ since $Y_i \perp Y_k |(Z_j, Z_m)$,} \\
     = &\ \{\hat Y_i-\pi_i(\btheta_M)\}\{\hat Y_k-\pi_k(\btheta_M)\}.
\end{align*}
Lastly, if $i\ne k$ but $i, k\in \mathcal{P}_j$, then the expectation of the $[i, k]$ entry of $\mathcal{D}(\btheta_M; \bY)$ given $\bZ$ is
\begin{align}
    &\ \mathbb{E}_{\hat \btheta_M}[\{Y_i-\pi_i(\btheta_M)\}\{Y_k-\pi_k(\btheta_M)\}|\bZ] \nonumber \\
     = &\ \mathbb{E}_{\hat \btheta_M}[Y_iY_k|Z_j]- \pi_k(\btheta_M)\hat Y_i- \pi_i(\btheta_M) \hat Y_k+\pi_i(\btheta_M)\pi_k(\btheta_M). \label{eq:lookinto}
\end{align}

To derive the expectation in \eqref{eq:lookinto}, we look into the probability $P_{\hat\btheta_M}(Y_iY_k=1|Z_j=z)$ for $z\in \{0, 1\}$ since $Y_iY_k$ is also a binary random quantity. Clearly, 
$$P_{\hat\btheta_M}(Y_iY_k=1|Z_j=z)=\frac{P_{\hat\btheta_M}(Y_i=1, Y_k=1, Z_j=z)}{P_{\hat \btheta_M}(Z_j=z)},$$
where, for the numerator,  
\begin{align*}
&\  P_{\hat\btheta_M}(Y_i=1, Y_k=1, Z_j=0) \\
= &\ P_{\hat\btheta_M}(Z_j=0|Y_i=1, Y_k=1)P_{\hat\btheta_M}(Y_i=1, Y_k=1) \\
= &\ (1-\text{Se})\pi_i(\hat\btheta_M)\pi_k(\hat\btheta_M), \text{ by the definition of sensitivity Se and due to $Y_i\perp Y_k$,}
\end{align*}
and, similarly, $P_{\hat\btheta_M}(Y_i=1, Y_k=1, Z_j=1)=\text{Se}\cdot \pi_i(\hat\btheta_M)\pi_k(\hat\btheta_M)$. Therefore, 
$$P_{\hat\btheta_M}(Y_iY_k=1|Z_j=z)=\frac{(1-\text{Se})^{1-z}\text{Se}^z \pi_i(\hat\btheta_M)\pi_k(\hat\btheta_M)}{\{P_{\hat \btheta_M}(Z_j=0)\}^{1-z}\{P_{\hat \btheta_M}(Z_j=1)\}^z},$$
which gives the expectation in \eqref{eq:lookinto}.

\setcounter{equation}{0} 
\def\theequation{B.\arabic{equation}}
\setcounter{figure}{0} 
\def\thefigure{B.\arabic{figure}}
\section*{Web Appendix B: A variance-covariance estimator}
\label{sec:B}
Following the variance estimation for an estimator resulting from an EM algorithm in \citet[][see Equation (2.9)]{elashoff2004algorithm}, we have the following sandwich form for estimating the pre-selection asymptotic variance of $\bar \btheta_M$, i.e., for estimating $\mathcal{I}^{-1}(\btheta_M)/n$, 
\begin{align}
    \mathcal{I}_c^{-1}(\hat \btheta_M;  \bZ) \left( \sum_{i=1}^n \left\{ \hat Y_i-\pi_i(\hat \btheta_M)\right\}^2  \begin{bmatrix}
        1 & \bx_{M,i}^\top \\
        \bx_{M,i} & \bx_{M,i} \bx^\top_{M,i} 
    \end{bmatrix}\right) \left\{\mathcal{I}_c^{-1}(\hat \btheta_M; \bZ)\right\}^\top, \label{eq:sandwich}
\end{align}
where $\mathcal{I}_c^{-1}(\hat \btheta_M;  \bZ)$ is the complete data information matrix in Equation (15) in the main article (that does not depend on unobserved true individual responses $\bY$) evaluated at $\hat \btheta_M$, that is, 
\begin{align*}
    \mathcal{I}_c(\hat \btheta_M;  \bZ) & = \sum_{i=1}^n \pi_i(\hat\btheta_M)\{1-\pi_i(\hat \btheta_M)\}
    \begin{bmatrix}
        1 & \bx^{\top}_{M,i} \\
        \bx_{M,i} & \bx_{M,i}\bx^{\top}_{M,i}
    \end{bmatrix}, 
\end{align*}
and, for $i=1, \ldots, n$, $\hat Y_i=\mathbb{E}_{\hat \btheta_M}[Y_i|Z_i]$ when individual testing data $\bZ=(Z_1, \ldots, Z_n)^\top$ are available, and $\hat Y_i=\mathbb{E}_{\hat \btheta_M}[Y_i|Z_j]$ when only group testing data $\bZ=(Z_1, \ldots, Z_J)^\top$ are available and $i\in \mathcal{P}_j$ for $j\in \{1, \ldots, J\}$. Extracting the lower $|M|\times |M|$ block of \eqref{eq:sandwich} gives an estimator for the variance of the pre-selection asymptotic distribution of $\bar \bbeta_M$.

\setcounter{equation}{0} 
\def\theequation{C.\arabic{equation}}
\setcounter{figure}{0} 
\def\thefigure{C.\arabic{figure}}
\setcounter{table}{0} 
\section*{Web Appendix C: Simulation results associated with the data splitting method}
Under the simulation setting described in Section 5.1 of the main article, we implemented the data splitting method based on group testing data. In particular, for each of 5000 Monte Carlo replicate dataset, we randomly divide $J$ pools into two subgroups, with  $J/2$ pools contributing data in the training set that we used to carry out variable selection following Algorithm 1 in the main article, and the testing set contains data from the remaining $J/2$ pools that we used to obtain point estimates and interval estimates of unknown quantities associated with the selected model. Table~\ref{table: splitting} provides summary statistics parallel to those in Tables 1 and 2 in the main article when $n=1000$ and $m=2$. 

Similar to results from the proposed post-selection procedure under the same setting, the empirical coverage probabilities for the selected covariate effects are close to the nominal level. The Monte Carlo average width of CIs for a null covariate effect resulting from the data splitting method is larger than that from the naive method but smaller than that from the proposed post-selection method.  

\begin{table}[]
\centering
\caption{\label{table: splitting}Monte Carlo averages of empirical coverage probabilities (C.P.) of $95\%$ CI's  for $\beta_2$, $\beta_4$ and $\beta_6$, along with the corresponding average confidence interval limits, average widths (A.W.) and average point estimates (P.E.). The results are based on a single value of $\lambda$ (the median AIC choice of $\lambda$) across 5000 Monte Carlo simulations at each combination of $n$ and $m$. Se=0.95, Sp=0.97. }
\begin{tabular}{cclccccc}
\hline
\multirow{2}{*}{$n$} & \multirow{2}{*}{$m$} & \multicolumn{1}{c}{\multirow{2}{*}{param}} & \multicolumn{5}{c}{Data Splitting}                      \\ \cline{4-8} 
                     &                      & \multicolumn{1}{c}{}                       & CI               & CI odds         & A.W. & C.P. & P.E. \\ \hline
                     & \multicolumn{1}{l}{} & $\beta_2$                                  & {[}0.20,2.40{]}  & {[}1.22,11.0{]} & 2.2  & 95.8 & 0.77 \\
1000                 & 2                    & $\beta_4$                                  & {[}-0.84,0.83{]} & {[}0.43,2.29{]} & 1.7  & 94.6 & 0.00 \\
                     & \multicolumn{1}{l}{} & $\beta_6$                                  & {[}-0.84,0.84{]} & {[}0.43,2.32{]} & 1.7  & 94.7 & 0.00 \\ \hline
\end{tabular}
\end{table}

\bibliographystyle{biom} \bibliography{ref}

\label{lastpage}